\documentstyle[12pt]{article}
\def\Journal#1#2#3#4{{\em #1} {\bf #2}, #3 (#4)}
\def\Book#1#2#3{{\em #1} ({#2}, {#3})}
\def\Preprint#1#2#3#4{{\em #1}, preprint~{#2}, {#3} (#4)}
\def\JPA{J. Phys. A: Math. Gen.}
\begin{document}
\newcommand{\be}{\begin{equation}}
\newcommand{\ee}{\end{equation}}
\newcommand{\bea}{\begin{eqnarray}}
\newcommand{\eea}{\end{eqnarray}}
\newcommand{\nn}{\nonumber}
\newcommand{\dbar}{\bar{\partial}}
\title{\bf Analytic-bilinear approach to integrable hierarchies.\\
II. Multicomponent KP and 2D Toda lattice hierarchies.}
\author{L.V. Bogdanov\thanks{
Permanent address: IINS, Landau Institute for Theoretical Physics, Kosygin
str. 2, Moscow 117940, GSP1, Russia; e-mail Leonid@landau.ac.ru}
\hspace{0.1em} and B.G. Konopelchenko\thanks{
Also: Budker Institute of Nuclear Physics, Novosibirsk 90, Russia} \\
Consortium EINSTEIN\thanks{
European Institute for Nonlinear Studies via Transnationally Extended
Interchanges}, \\
Dipartimento di Fisica dell'Universit\`a di Lecce \\and Sezione INFN,
73100 Lecce, Italy}
\date{}
\maketitle
\begin{abstract}
Analytic-bilinear approach for construction and study of integrable
hierarchies is discussed.
Generalized multicomponent KP and 2D Toda lattice hierarchies
are considered. This approach allows to represent
generalized hierarchies of integrable equations in a condensed form of
finite functional equations.
Generalized hierarchy incorporates basic hierarchy,
modified hierarchy, singularity manifold equation hierarchy
and corresponding linear problems.
Different levels of generalized hierarchy are connected
via invariants of Combescure symmetry transformation.
Resolution of functional equations also leads
to the $\tau $-function and addition formulae to it.
\end{abstract}
\section{Introduction}
The Kadomtsev-Petviashvili (KP) equation and the whole KP hierarchy as well
as the multicomponent KP and
two-dimensional Toda lattice hierarchies are among
the most important ingredients of the theory of integrable partial
differential equations. They arise in various fields of physics from
hydrodynamics to string theory.

Integrable hierarchies have been described and studied
within the framework of different approaches. The Sato approach
\cite{Sato}-\cite{Segal}
and the $\bar{\partial}$-dressing method (see \cite{dbar1}-
\cite{dbar3} and review \cite{dbar4})
are among the most beautiful and powerful among them. These two
approaches look like completely different. The Sato approach is
in essence algebraic  with its infinite-dimensional Grassmannian,
pseudo-differential operators, Hirota bilinear identity
and $\tau$-function. In contrast, the $\bar{\partial}$-dressing method,
based on the nonlocal $\bar{\partial}$-problem, mostly uses the
technique of the complex variable theory.

Recently we proposed (\cite{Comb},
\cite{KP1}) an approach which combines
the characteristic features of both these methods,
namely the Hirota bilinear identity from the Sato approach
and analytic properties of the wavefunctions
from the $\bar{\partial}$-dressing method.
This analytic-bilinear approach is based on the
generalized Hirota bilinear identity
for the wave function $\psi(\lambda,\mu)$ with simple analytic
properties (Cauchy-Baker-Akhiezer (CBA) function).
The CBA function (or, to be more precise,
its `nonsingular part' $\chi(\lambda,\mu)$ is
defined in some region $G$ of the complex plane
and it is characterized by the generalized Hirota
bilinear identity on the boundary of this region
\cite{NLS}. The CBA function arises both
in $\bar{\partial}$-dressing method
\cite{dbar2} and in finite gap integration method
\cite{Orlov}, and it can be introduced also to the context
of general Grassmannian approach \cite{Segal},
\cite{Sato}.

According to generalized Hirota
bilinear identity, the CBA function is a functional
of the function $g$ defined on the boundary of $G$
(in fact, $g$ is a map of $\partial G$
to {\bf C}$^{\times}$ (i.e. {\bf C}-0), or,
in other words,
it is an element of loop group).
One can choose some
special subgroups of the generic loop group
parametrized by finite number of variables
(`times') to get particular examples of
integrable equations. This is a standard approach
in the $\bar{\partial}$-dressing method. One may also
try to use some generic loop group and extract a
full infinite set of equations. So the notion of the hierarchy
arises. The dependence of the CBA function on $g$ encodes
the dynamics of the system.
The hierarchy can be written down in variational form,
i.e. as equations describing variations of the objects of theory
under the arbitrary variation of group element $g$. To get
an hierarchy in terms of times, one should introduce
a parametrization of the loop group.

The distinctive feature of analytic-bilinear approach
is that our primary object is the CBA function and
hierarchies are written in the form
of equations for the CBA function (we call these equations
{\em generalized} hierarchies).
The analytic-bilinear approach allows us to describe
generalized hierarchies in a compact finite form.
Such a generalized hierarchy contains basic integrable hierarchy,
hierarchy of equation for wave functions
(modified hierarchy)
and hierarchy of equations for CBA-type wave functions
(we call it singularity manifold equations (SM) hierarchy,
because in the scalar KP case it coincides with the
singularity manifold equations obtained from the Painleve
analysis of KP; however, our contexts is completely different),
and corresponding linear problems.
The relation between three hierarchies of equations
can be described in terms of invariants of
Combescure symmetry group (\cite{Comb}, \cite{KP1}).

The generalized Hirota bilinear identity
provides various functional equations for the CBA function.
The resolution of some of them leads to the introduction
of the $\tau$-function, while others give rise to the addition
theorems for the $\tau$-function.
The properties of the $\tau$-function,
such as the associated closed 1-form and its global definition are also
arise in a simple manner within the analytic-bilinear approach.

The first part of this paper \cite{KP1}
was devoted to the scalar KP hierarchy.
In this case $G$ is just a domain of the complex plane.
Without the loss of generality, we choose $G$ as unit disc.
The group element $g$ is defined on the unit circle
and can be represented in a form
$$
g(\lambda)=\lambda^n g_{\rm in}\times g_{\rm out}
$$
where $n$ is the index
of the function, $g_{\rm in}$ and $g_{\rm out}$
are functions analytic and
having no zeroes respectively inside
(outside) the unit circle on the complex plane
(i.e. elements of the loop groups $\Gamma_-$ and
$\Gamma_+$).
In the scalar KP hierarchy case, we have no opportunity
to treat index variable without some additional construction,
so we put index to zero. The functions $g_{\rm out}$ and $g_{\rm in}$
are parametrized in a standard way  by infinite sets of
`positive' and `negative' times. But it appears
that in the scalar KP hierarchy case dependence of the objects
of theory on $g_{\rm in}$ (i.e. on negative times in
our notations) is trivial,
so the scalar KP hierarchy is defined effectively
by the subgroup $\Gamma_+$ of the loop group $\Gamma$.

In the present second part of the paper we consider the
generalized multi-component KP
hierarchy and the two-dimensional Toda lattice hierarchy
(for other approaches to the usual multicomponent KP
hierarchy and 2D Toda lattice hierarchy see
\cite{Sato}, \cite{Jimbo}, \cite{Toda1}, \cite{Toda2},
\cite{Kac}, \cite{Dickey}).

There are two equivalent approaches to investigate multi-component
hierarchy. One could start with the generalized Hirota identity for
the matrix-valued
CBA function. In this case $G$ is again a unit disc,
and $g$ is a diagonal matrix function on the unit circle
(i.e. a set of $N$ functions $g_n$). Another way is to
start with the scalar CBA function defined in the
region $G$ consisting of $N$ disconnected domains
(unit discs for simplicity). In this case
the dynamics is also defined by the set of $N$ functions
$g_n$, and the CBA function consists, in fact,
of $N\times N$ disconnected components.
These two representations are exactly equivalent
and are associated with the
multicomponent loop group $\Gamma^N$.
In our paper we will use both of them.

As in the scalar case, the objects of multicomponent
KP theory depend trivially on the subgroup $\Gamma_-^N$.
The difference with the scalar case consists in the presence
of discrete variables (indices), though only the lattice
vectors with zero general index are admissible.

Since the generalized Hirota bilinear identity is
our primary object,
it is quite natural to proceed
from the hierarchy corresponding to the simplest choice of
$G$ -- the unit disk -- to the hierarchies corresponding
to $G$ consisting of several unit discs, i.e. from
the scalar KP hierarchy to multicomponent KP hierarchy.
But on the other hand, the loop group is also
a fundamental object of the theory,
so one would like to find an hierarchy with a full-scale
action of generic loop group on the objects of theory.
This hierarchy is known as Toda-lattice hierarchy.
In our approach it can be introduced as a two-component
KP hierarchy with the special choice of $G$ consisting
of two discs with the common boundary (the unit circle):
one disc around origin and one disc around infinity.
In this case, up to some trivial dynamics, one can glue
two special loops defining the two-component KP hierarchy
to obtain a generic loop. In the case of Toda lattice
hierarchy the discrete index variable, positive and negative
time variables get their natural interpretation.

We first consider the multicomponent KP hierarchy.
We start with the generalized Hirota identity for the matrix-valued
CBA function and derive various finite functional equations
for this function. Part of these functional equations present the compact
form of the generalized multicomponent KP hierarchy which includes
equations for potentials
(the standard multicomponent KP hierarchy),
equations for the wave functions
(the modified multicomponent
KP hierarchy),
equations for CBA-type wave functions
(singularity manifold equation hierarchy)
and corresponding linear problems.

Resolution of functional equations introduces the set of
$\tau$-functions. We discuss also the associated closed 1-form
and its global definition. The simplest integrable systems
are given by the resonantly interacting wave equations and
by the scalar Darboux system which is its modified partner.

We present a compact form of {\em equations} of multicomponent
KP hierarchy, modified KP hierarchy and singularity manifold
equation hierarchy, using shift operators. Similar formulae
for the first two cases can be found in \cite{Nijhoff}.

The two-component case which corresponds to the generalized
Davey-Stewartson (DS) hierarchy is considered in detail.

The description of the generalized Toda lattice hierarchy within the
analytic-bilinear approach requires minor modifications of the
multicomponent KP case. For simplicity we consider here only
the Abelian Toda lattice hierarchy. It appears that
modified Toda lattice equation (i.e. equation for the
Toda lattice wave functions) coincides with the
2D Volterra system, which was recently discussed in literature
in different contexts \cite{Fer}, \cite{Shabat}. We
derive 2D Toda lattice singularity manifold equation,
which was also discussed in \cite{Gibbon}.
The invariants of Combescure transformations
group (\cite{Comb}, \cite{KP1}) are found.
We also present the hierarchies of equations using
shift operators.

The paper is organized as follows.
The generalized Hirota bilinear identity and some basic
formulae are discussed in section 2.
Generalized multicomponent KP
hierarchy is considered in section 3.
Compact forms of equations of generalized multicomponent KP
hierarchy are presented in section 4.
Section 5 is devoted to generalized DS hierarchy.
The $\tau$-function for the DS hierarchy
is introduced in section 6.
The $\tau$-function and corresponding addition formulae for
the multicomponent KP hierarchy
are considered in section 7.
Closed variational 1-form and global definition for the
$\tau$-function are given in section 8.
Basic equations for the generalized 2D Toda lattice hierarchy
are derived in section 9. Concrete equations associated with
the generalized 2DTL hierarchy are presented in section 10.
Addition formulae for the 2DTL hierarchy are given in
section 11.
\section{Generalized Hirota identity, functional equations
and generalized multicomponent KP hierarchy}
The analytic-bilinear approach starts with the generalized
Hirota bilinear identity
\be
\int_{\partial G} \chi(\nu,\mu;g_1)g_1(\nu)g_2^{-1}(\nu)
\chi(\lambda,\nu;g_2)d\nu=0,
\label{HIROTA}
\ee
where $\chi(\lambda,\mu;g)$ is the CBA function
normalized by the condition $\chi\rightarrow (\lambda-\mu)^{-1}$
as $\lambda
\rightarrow\mu$ and
$\chi(\lambda,\mu)$ is analytic function of
both variables $\lambda,\mu$ for $\lambda\neq\mu$
in the domain $G$ with the boundary $\partial G$.
The quantity $g=g(\lambda,{\bf x})$ specifies the dependence
on the infinite set of dynamical variables
${\bf x}=(x_1\,,x_2\,,...)$.
In terms of the wavefunction $\psi(\lambda,\mu;g)$
defined via
\be
\psi(\lambda,\mu;g)=g^{-1}(\mu)\chi(\lambda,\mu;g)g(\lambda)
\label{substitution}
\ee
the generalized Hirota bilinear identity looks like
\be
\int_{\partial G} \psi(\nu,\mu;g_1)
\psi(\lambda,\nu;g_2)d\nu=0.
\label{HIROTA1}
\ee
Note that the generalized Hirota bilinear identity
is bilocal both in dynamical and spectral variables.

In this paper we consider the case for which $\chi$
and $\psi$ take their value in the full linear group
GL(N,{\bf C}), $G$ is a unit disc with a center $\lambda=0$.
Elements $g(\lambda,{\bf x})$ belong to
maximal abelian subgroup of GL(N,{\bf C}).

It is important to note that there is an alternative
way to treat the matrix problem (\ref{HIROTA}),
starting from the {\em scalar} identity but
with the region $G$ consisting of several
disconnected domains.
In this case one has to choose
domain $G$ consisting of N unit
discs $D_\alpha$ with the center at $\lambda=0$.
Now the scalar CBA function effectively
consists of $N\times N$ disconnected components and
can be interpreted as matrix function defined
on the unit disc:
\be
\chi_{\alpha\beta}(\lambda,\mu)=
\chi(\lambda,\mu),\;\lambda\in D_\beta,\,
\mu\in D_\alpha.
\label{matrix}
\ee
The function $g$ in this framework is a set
of $N$ functions defined on the unit circle.
Thus, rewriting the scalar identity in matrix
notations, we get {\em exactly} the same
matrix generalized Hirota identity,
where $g$ is a diagonal matrix.

So the matrix
structure in the scalar case is hidden in the
multicomponent domain of definition of the CBA
function. The advantage of the scalar treatment is
the opportunity to preserve most of the structure
of the one-component KP hierarchy case almost
intact and
to use geometric intuition in the matrix case.
We will use both representations
of the generalized Hirota identity in the paper.

The dependence on dynamical variables is hidden
in the set of $N$ functions $g$ defined on the
unit circle. The most natural choice would be
to treat sets of arbitrary functions on the
circle having no zeroes, forming the group
by multiplication.
Arbitrary
$g$ can be represented in the form
$$
g(\lambda)=\lambda^n g_{\rm in}\times g_{\rm out}
$$
where discrete variable $n$ is the index
of the function, $g_{\rm in}$ and $g_{\rm out}$
are functions analytic and
having no zeroes respectively inside
(outside) the unit circle on the complex plane.
We will use the following parametrization
for $g_{\rm in}$ and $g_{\rm out}$:
\be
g_{\alpha\;{\rm out}}=
\exp\left(\sum_{n=1}^{\infty}x_{(\alpha) n} \lambda^{-n})
\right),
\label{out}
\ee
\be
g_{\alpha\;{\rm in}}=
\exp\left(\sum_{-\infty}^{n=-1}x_{(\alpha) n} \lambda^{-n})
\right).
\label{in}
\ee
In these formulae we have omitted `the zero time'
$x_0$, that defines simply some constant factor
in front of $g$ and, according to (\ref{HIROTA}),
it doesn't change the function $\chi$.

So, in the general case the set of functions
$g_\alpha$ can be parametrized by $N$ discrete
times $n_{\alpha}$ and $N$ infinite sets of
continuous variables $x_{\alpha n}$, $n\in{\bf Z}$.
But in the case of KP hierarchy the dependence
of the function $\chi(\lambda,\mu;g)$ on
$g_{\rm in}$ (in other words,
on the `negative times') is trivial; according to
(\ref{HIROTA}) (the `scalar' form),
we get
\be
\chi(\lambda,\mu;g\times g_{\rm in})=
g_{\rm in}^{-1}(\mu)\chi(\lambda,\mu;g)g_{\rm in}(\lambda).
\ee
It occurs that other important objects (KP solution,
$\tau$-function) also depend on negative times
trivially. Thus, for simplicity, from the very beginning we omit
the dependence on the negative times.

As for discrete times, we will see, that
the case of multicomponent KP hierarchy allows
only sets of discrete times with the zero
general index $(\sum_{\alpha=1}^{N}n_{\alpha}=0)$.

So the case of multicomponent KP hierarchy seems
a bit asymmetric with respect to positive and negative
times, and it also doesn't allow to consider arbitrary
lattice vectors of discrete times. The most natural
way to resolve these problems is to consider
multicomponent Toda-lattice hierarchy (see below).
The objects of Toda-lattice hierarchy have
a natural dependence on positive, negative and
discrete times.

To find special solutions for (\ref{HIROTA}), one may
use some additional construction
(e.g. $\bar{\partial}$-dressing method or
algebro-geometric technique). But it is possible
to develop a procedure of solving (\ref{HIROTA})
in terms of (\ref{HIROTA}) itself. In this case
the function $\chi(\lambda,\mu,g=1)=\chi_0(\lambda,\mu)$
should be treated as initial data, and one can start
from arbitrary $\chi_0(\lambda,\mu)$ possessing the
necessary analytical properties. Then the dynamics
of the function $\chi(\lambda,\mu)$ can be defined
via the variational form of identity (\ref{HIROTA})
\bea
\delta \chi(\lambda,\mu;g)=
\oint \chi(\nu,\mu;g){\delta g(\nu)\over g(\nu)}
\chi(\lambda,\nu;g)d\nu\,.
\label{variation}
\eea
So the problem of solving (\ref{HIROTA}) is reduced
to (infinite number in terms of variables $x_n$)
differential equations with initial condition.
Taking the second variation in $\delta(\ln g)$,
it is easy to prove that
this system is self-consistent (i.e. the
cross-derivatives are equal), and it is a correctly
defined variational equation for the function
$\chi(\lambda,\mu,g)$. Going further, one could
take higher variational derivatives (all of them have
a very simple form) and represent a formal solution
in terms of Taylor-type series.

Another approach to solving  (\ref{HIROTA})
is to extract from it some linear integral
equations. Indeed, let us choose some initial
condition $\chi_0(\lambda,\mu)$. Then for the
function $\chi(\lambda,\mu,g)$ from
(\ref{HIROTA}) we have a pair of equations
\be
\int_{\partial G} \chi(\nu,\mu;g)g(\nu)
\chi_0(\lambda,\nu)d\nu=0,
\label{HER}
\ee
\be
\int_{\partial G} \chi_0(\nu,\mu)g^{-1}(\nu)
\chi(\lambda,\nu;g)d\nu=0.
\label{HIR}
\ee
These relations can be interpreted as linear integral
equations for the function $\chi(\lambda,\mu;g)$;
integral equations of such type arise in the context of
nonlocal Riemann problem. Using the technique
developed in \cite{NLS}, one may show that if
both of these equations are solvable (in
the class of functions $\chi(\lambda,\mu)$ with the analytical
properties specified above, here they play a role of
{\em normalization}),
then the
solution for both equations is the same and it is unique.

There is a very important class of functions $g$
corresponding to some special discrete transformations
-- meromorphyc functions -- for which the dynamics can
be found explicitly. Namely,the solution for both
equations (\ref{HER}), (\ref{HIR}) exists
(and  it is unique),
if and only
if meromorphic function $g$ has equal number of
zeroes and poles in $G$ (taking into account
the multiplicity),
and it is given by the
determinant formula \cite{NLS}. This statement
means that the dependence of $\chi$ on
discrete (index) variables is correctly
defined for the {\em zero general index case} and,
moreover, it can be represented in explicit form
\cite{NLS}.
\section{Generalized multicomponent KP hierarchy}
Proceeding with the multicomponent KP case, we choose
the following parametrization for $g_{\rm out}$
(matrix representation)
\be
g(\lambda,{\bf x})=\exp\left[\sum_{\alpha=1}^N\sum_{n=1}^{\infty}
{E_{\alpha}\over \lambda^n}x_{(\alpha)\,n}\right],
\label{g}
\ee
where the matrices $E_{\alpha}$ form a basis of the commutative
subalgebra
of diagonal matrices: $(E_{\alpha})_{\beta\,\gamma}=
\delta_{\alpha\beta}\delta_{\beta\gamma}\quad
(\alpha,\beta,\gamma=1,...,N)$.
So we have $N$ infinite series of dynamical variables
$x_{(\alpha)\,n}$.

Equations (\ref{HIROTA}) or (\ref{HIROTA1}) with $g$ of the form
(\ref{g}) generically define the multicomponent KP hierarchy.
To extract from them finite functional equations
similar to the scalar case we consider the Hirota identity
(\ref{HIROTA}) for the dynamical variables $x'_{(\alpha)\,n}$
and $x_{(\alpha)\,n}$related in a special way, namely
\be
x'_{(\alpha)\,n}-x_{(\alpha)\,n}={1\over n}a_{\alpha}^n-
{1\over n}b_{\alpha}^n\,,
\ee
where $a_{\alpha}$, $b_{\alpha}$ ($0\geq\alpha\geq N$)
are arbitrary parameters. We will also use vector notations
${\bf x}=({\bf x_1},...,{\bf x_N})$,
${\bf x_\alpha}=(x_{(\alpha)1},x_{(\alpha)2},...)$.
In this terms the preceding formula looks like
\be
{\bf x'_{(\alpha)}}-{\bf x_{(\alpha)}}=[a_{\alpha}]-
[b_{\alpha}]\,,
\ee
or
\be
{\bf x'}-{\bf x}=[{\bf a}]-
[b]\,,
\ee
where $[{\bf a}]_{(\alpha)i} ={\frac{1}{i}} a_\alpha^i$.
For such $x$ and $x'$ due to the formula
$\log(1-\epsilon)=\sum_{n=1}^{\infty}{\epsilon^n\over n}$
one has
\be
\left(g_1(\nu)g_2^{-1}(\nu)\right)_{\alpha\,\beta}=
\delta_{\alpha\,\beta}{\nu-a_{\alpha}\over\nu-b_{\alpha}}
\quad 0\geq\alpha\,,\beta\geq N.
\label{ratio}
\ee
Substitution of the expression (\ref{ratio}) into (\ref{HIROTA})
gives
\begin{eqnarray}
&&\left({\frac{\mu-a_{\alpha}}{\mu-b_{\alpha}}}\right)
\chi_{\alpha\beta}(\lambda,\mu,{\bf x}+[{\bf a}])- \left({
\frac{\lambda-a_{\beta}}{\lambda-b_{\beta}}}\right)
\chi_{\alpha\beta}(\lambda,\mu,{\bf x}+[{\bf b}])+  \nonumber
\\
\nonumber \\
&&
\sum_{\gamma=1}^{N}
(b_{\gamma}-a_{\gamma})
\chi_{\gamma\,\beta}(\lambda,b_{\gamma},{\bf x}+[{\bf a}])
\chi_{\alpha\,\gamma}(b_{\gamma},\mu,{\bf x}+[{\bf b}])=0.
\label{KPbasic0}
\end{eqnarray}
Equation (\ref{KPbasic0}) represents the final
functional form of the generalized multicomponent KP hierarchy.
It encodes the usual multicomponent KP hierarchy and
its modified version together with their linear problems.
To demonstrate this fact we, similar to the scalar case
\cite{KP1}, consider the particular form of equation (\ref{KPbasic0})
with $b_{\alpha}=0, \quad0\geq\alpha\geq N$
\begin{eqnarray}
&&\left(1-{a_\alpha\over\mu}\right)
\chi_{\alpha\beta}(\lambda,\mu,{\bf x}+[{\bf a}])- \left(
1-{a_\beta\over\lambda}\right)
\chi_{\alpha\beta}(\lambda,\mu,{\bf x})=  \nonumber
\\
\nonumber \\
&&
\sum_{\gamma=1}^{N}
a_{\gamma}
\chi_{\gamma\beta}(\lambda,0,{\bf x}+[{\bf a}])
\chi_{\alpha\gamma}(0,\mu,{\bf x})=0,
\label{KPbasic01}
\end{eqnarray}
or, in terms of the CBA and Baker-Akhiezer functions
\begin{eqnarray}
&&
\psi_{\alpha\beta}(\lambda,\mu,{\bf x}+[{\bf a}])-
\psi_{\alpha\beta}(\lambda,\mu,{\bf x})=
\sum_{\gamma=1}^{N}
a_{\gamma}
\psi_{\gamma\,\beta}(\lambda,{\bf x}+[{\bf a}])
\widetilde\psi_{\alpha\gamma}(\mu,{\bf x})
\label{KPbasic02}
\end{eqnarray}
where $\psi_{\gamma\,\beta}(\lambda,{\bf x})=
\chi(\lambda,0,{\bf x})g_\beta(\lambda)$,
$\widetilde\psi_{\gamma\,\beta}(\mu,{\bf x})=
\chi(0,\mu,{\bf x})g_\gamma^{-1}(\mu)$.
The expansion of (\ref{KPbasic02}) in
Taylor series in $a_{(\gamma)}$ gives rise in the first order
to the following equation
\be
{\partial \psi_{\alpha\beta}(\lambda,\mu,{\bf x})
\over
\partial x_{(\gamma)\,1}
}
= \psi_{\gamma\beta}(\lambda,{\bf x})
\widetilde\psi_{\alpha\gamma}
(\mu,{\bf x})\,.
\label{KPbasic11}
\ee
Evaluating the equation (\ref{KPbasic01})
with $\alpha\neq\beta\neq\gamma\neq\alpha$
at $\lambda=\mu=0$, one obtains the system
\be
{\partial P_{\alpha\beta}\over\partial \xi_{\gamma}}=
P_{\alpha\gamma}P_{\gamma\beta}\,,
\label{N-wave}
\ee
where $P_{\alpha\beta}=\chi_{\alpha\,\beta}(0,0,{\bf x})$,
$\xi_{\gamma}=x_{(\gamma)1}$,
which is nothing but the Darboux system in terms
of rotation coefficients \cite{Darboux} or resonantly
interacting waves equation \cite{dbar1}.
Equation (\ref{KPbasic01}) with $\alpha\neq\gamma$
and $\mu=0$ gives the system
\be
{\partial \Psi_{\alpha}\over\partial \xi_{\gamma}}=
P_{\alpha\gamma}\Psi_{\gamma}\,,
\label{linear}
\ee
where
\be
\Psi_{\alpha}=\sum_{\beta=1}^{N}\oint
d\lambda f_\beta(\lambda)\psi_{\alpha\beta}(\lambda,{\bf x})
\label{Psi}
\ee
and $f_\beta(\lambda)$ are arbitrary functions. The system
(\ref{linear}) is a well-known linear problem for equations
(\ref{N-wave}).

Equation (\ref{KPbasic01}) with $\beta\neq\gamma$
and $\lambda=0$ gives the adjoint linear problem for the
system (\ref{N-wave})
\be
{\partial \widetilde\Psi_{\beta}\over\partial \xi_{\gamma}}=
P_{\alpha\gamma}\widetilde\Psi_{\gamma}\,,
\label{linear+}
\ee
where $\widetilde\Psi_{\beta}=\sum_{\alpha=1}^{N}\oint
d\mu \widetilde f_\alpha(\mu)
\widetilde\psi_{\alpha\beta}(\mu,{\bf x})$
and $\widetilde f_\alpha(\mu)$ are arbitrary functions.

Further (\ref{linear}) implies that
$P_{\alpha\gamma}={1\over\Psi_\gamma}{\partial \Psi_\alpha
\over \partial \xi_\gamma}$.
Substituting this expression into (\ref{N-wave}),
one gets
\bea
\partial_\beta\partial_\gamma\Psi_\alpha&=&
{\Psi_\beta}^{-1}(\partial_\gamma\Psi_\beta)
(\partial_\beta \Psi_\alpha)
+{\Psi_\gamma}^{-1} (\partial_\beta\Psi_\gamma)
(\partial_\gamma
\Psi_\alpha),
\label{Darboux}
\eea
where $\partial_\beta={\partial\over\partial\xi_\beta}$.
This is a well-known Darboux-Zakharov-Manakov system
which has been discovered in the theory of systems of
conjugate surfaces first \cite{Darboux} and then has been
rediscovered within the $\bar{\partial}$-dressing method
in \cite{dbar1}. This system is a modified partner of the
system (\ref{N-wave}). To get the linear problem for it,
it is sufficient to substitute the expression
$P_{\alpha\gamma}={1\over\Psi_\gamma}{\partial \Psi_\alpha
\over \partial \xi_\gamma}$
into (\ref{linear+}).

Thus the system (\ref{KPbasic01}) contains the equation
(\ref{N-wave}), its modified partner (\ref{Darboux})
and their linear problems.

The higher order terms in the expansion of (\ref{KPbasic01})
in $a_\gamma$
give rise to the higher analogues of equation (\ref{KPbasic11}).
It is a straightforward check that these equations contain the
higher order equations (\ref{N-wave}), their modified analogues
and corresponding linear problems.

The generalized multicomponent KP hierarchy described by
functional equation (\ref{KPbasic0}) has rather special
structure. In particular, it contains N independent
generalized scalar KP hierarchies. They are obviously given
by equations (\ref{KPbasic0}) with $\alpha=\beta$
and $b_{\gamma}=a_{\gamma}$ for all $\gamma$
except $\gamma=\alpha$. It is also quite clear that the number of
effective independent dynamical variables is equal to three.
For instance, the systems (\ref{N-wave}) and
(\ref{Darboux}) are overdetermined and represent a collection
of commuting closed systems defined for every triad of dynamical
variables.

Note also that one can get the multicomponent KP hierarchy
starting with the scalar generalized Hirota identity
(\ref{HIROTA}). In this case one has to choose
disconnected domain $G$ consisting of N unit
discs $D_\alpha$ with the center at $\lambda=0$.
In this case the scalar CBA function effectively
consists of $N\times N$ disconnected components and
can be interpreted as matrix function. So the matrix
structure in the scalar case is hidden in the
multicomponent domain of definition of the CBA
function. The function $g(\lambda)$
is defined on the set of $N$ unit circles
$S_\alpha$ and consists in fact of
$N$ components $g_\alpha$ defined on a unit circle.

The advantage of the scalar treatment is
the opportunity to preserve most of the structure
of the one-component KP hierarchy case almost
without modification.

Using the parametrization (\ref{out}),
we represent the function $g(\lambda)$
(defined on the set of $N$ unit circles
$S_\alpha$) as
\be
g({\bf x},\lambda)
=\prod_{\alpha=1}^{N}
\exp\left(\sum_{i=1}^{\infty}x_{(\alpha)\,i}K_{\alpha}^i\right)\,,
\label{gKP}
\ee
where
\bea
K_\alpha(\lambda)={1\over \lambda},\quad \lambda\in{\bf D}_\alpha;\nn\\
K_\alpha(\lambda)=0,\quad \lambda\notin {\bf D}_\alpha.\nn
\eea
In this case the transformation group element $g_0$
providing just the same shift of the ${\bf x}$ variables
takes the form
\be
g_0(\nu)=\left(g_1(\nu)g_2^{-1}(\nu)\right)=
{\nu-a_{\alpha}\over\nu-b_{\alpha}}
\quad \nu\in{\bf D}_\alpha\,.
\label{ratio1}
\ee
The analogue of equation (\ref{KPbasic0}) reads
\begin{eqnarray}
&&g_0(\mu)\chi(\lambda,\mu,{\bf x}+[{\bf a}])-
g_0(\lambda)\chi(\lambda,\mu,{\bf x}+[{\bf b}])+
\nonumber \\
&&
\sum_{\gamma=1}^{N}
(b_{\gamma}-a_{\gamma})
\chi(\lambda,b_{(\gamma)},{\bf x}+[{\bf a}])
\chi(b_{\gamma},\mu,{\bf x}+[{\bf b}])=0.
\label{KPbasic2}
\end{eqnarray}
It seems to be a scalar equation, but if we take into account
that the function $\chi$ consists of a set of disconnected components
and denote $\chi_{\alpha\beta}(\lambda,\mu)=\chi(\lambda,\mu),
\quad \lambda\in D_{\beta},\mu\in D_{\alpha}$, then this equation
gives rise exactly to the system (\ref{KPbasic0}).

The analogue of equation (\ref{KPbasic01}) is
respectively
\begin{eqnarray}
&&\prod_{\alpha=1}^{N}(1-a_\alpha K_\alpha(\mu))
\chi(\lambda,\mu,{\bf x}+[{\bf a}])-
\prod_{\alpha=1}^{N}(1-a_\alpha K_\alpha(\lambda))
\chi(\lambda,\mu,{\bf x})=
\nonumber \\
&&
\sum_{\gamma=1}^{N}
a_{(\gamma)}
\chi_\gamma(\lambda,{\bf x}+[{\bf a}])
\widetilde\chi_\gamma(\mu,{\bf x})=0,
\label{KPbasic21}
\end{eqnarray}
and in terms of CBA and BA functions one gets
\begin{eqnarray}
&&\psi(\lambda,\mu,{\bf x}+[{\bf a}])-
\psi(\lambda,\mu,{\bf x})=
\sum_{\gamma=1}^{N}
a_{\gamma}
\psi_{\gamma}(\lambda,{\bf x}+[{\bf a}])
\widetilde\psi_{\gamma}(\mu,{\bf x}),
\label{KPbasic22}
\eea
where we have introduced notations
\bea
&&
\chi_\gamma(\lambda,{\bf x})=
\chi(\lambda,0_\gamma,{\bf x}),
\nn\\&&
\widetilde\chi_\gamma(\mu,{\bf x})=
\chi(0_\gamma,\mu,{\bf x}),
\label{chi}\\
&&
\psi_\alpha(\lambda,{\bf x})=g(\lambda)\chi_\alpha(\lambda,{\bf x}),
\nn\\&&
\widetilde\psi_\alpha(\mu,{\bf x})
=g^{-1}(\mu)\widetilde\chi_\alpha(\mu,{\bf x}),
\nn
\eea
$0_\alpha$ denotes the center point of
the disc $D_\alpha$.

We will also use following notations, most of
them exactly corresponding to notations introduced
in frame of matrix representation:\\
for potentials
$$
\chi_{\alpha\beta}({\bf x})=\chi_{\rm r}(0_\beta,0_\alpha,{\bf x}),
$$
$$
\chi_{\rm r}(\lambda,\mu,{\bf x})=
\chi(\lambda,\mu,{\bf x})-\delta_{\alpha\beta}{1\over\lambda-\mu},
\quad
\lambda\in D_\alpha,\;\mu\in D_\beta
$$
for wave functions
\bea
&&
\Phi({\bf x})=\int\!\!\!\int_{\partial G}
d\lambda d\mu f(\lambda) \widetilde f(\mu)
\psi(\lambda,\mu,{\bf x}),
\nn\\&&
\Psi_\alpha({\bf x})=\int_{\partial G}
d\lambda f(\lambda)
\psi_\alpha(\lambda,{\bf x}),
\nn\\&&
\widetilde\Psi_\alpha({\bf x})=\int_{\partial G}
d\mu \widetilde f(\mu)
\widetilde
\psi_\alpha(\mu,{\bf x}),
\label{Phi}
\eea
where $f(\lambda), \widetilde f(\mu)$ are arbitrary functions
defined on the boundary of $G$ (i.e. on $N$ unit circles).
\section{Equations
of generalized KP hierarchy in compact form}
The basic equation of the multicomponent KP hierarchy
(\ref{KPbasic01}) looks very simple and contains all the information
about generalized hierarchy -- hierarchy of linear problems
and hierarchy of equations. However, sometimes this information
could be excessive and one would prefer to have a compact
form of the {\em equations} of the hierarchy. As we have demonstrated
in the previous section in the case of DS hierarchy,
it is possible to extract all the hierarchy of equations
from (\ref{KPbasic01}) in terms of shift operators.
To derive the equations of the KP hierarchy,
we will use generic shifts of times. For each triad
of different shifts (i.e. shifts with different parameters)
it is possible to derive a closed system of equations,
representing, in fact, the multicomponent KP hierarchy.

As we will see later, the multicomponent KP hierarchy
in a general case incorporates also discrete times not reducible
to the shifts of continuous times. So equations of
multicomponent KP hierarchy can also contain
these essentially discrete times. It is possible
to include them to equations of this section,
but now we will derive the equations of KP hierarchy
containing only continuous times.
Discrete times will be discussed later in the context
of Toda lattice hierarchy.

Thus, we
will derive a compact form of the equation of multicomponent
KP hierarchy, modified KP hierarchy and singularity manifold
equation hierarchy. It is interesting to note that these
equations are obtained in fact in the form of lattice
equations, and all the hierarchy is given by the expansion
of these lattice equations with respect to the
parameters of the lattice.

In this section we will use the matrix form
of the equation  (\ref{KPbasic01}), i.e.
\bea
&&
\Delta\chi(\lambda,\mu)+
{1\over\lambda}\chi(\lambda,\mu)A-
{1\over\mu}A\cdot T\chi(\lambda,\mu)=
\chi(0,\mu)A\,T\chi(\lambda,0).
\label{KPB}
\eea
Here the common argument ${\bf x}$ is omitted,
the function
$\chi(\lambda,\mu,{\bf x})$ is
$N\times N$ matrix-valued function,
$A$ is a diagonal matrix, $A_{ij}=\delta_{ij}a_i$,
and we have introduced shift operator
and difference operator
\bea
&&
T\chi(\lambda,\mu,{\bf x})=\chi(\lambda,\mu,{\bf x}+[{\bf a}]),
\nn\\&&
\Delta\chi(\lambda,\mu,{\bf x})=(T-1)\chi(\lambda,\mu,{\bf x}).
\nn
\eea
If one takes the first term of expansion of the relation
(\ref{KPB}) at $\lambda=0$, $\mu=0$, one obtains
an identity containing the potential
$u=\chi_{\rm r}(0,0,{\bf x})$ and two higher terms of
expansion of $\chi_{\rm r}(\lambda,\mu,{\bf x})$.
It is possible to cancel these terms by combining
three equations (\ref{KPB}) with different shift
parameters ${\bf a_k},\; k\in \{1,2,3\}$ and get an
equation for the potential $u$.

First, it is easy to cancel the terms containing $A_i/\lambda$
combining pairs of equations
\bea
&&
\Delta_i \chi(\lambda,\mu)A_j-
\Delta_j \chi(\lambda,\mu)A_i+{1\over\mu}
(A_iT_i\chi(\lambda,\mu)A_j- A_jT_j\chi(\lambda,\mu)A_i)=
\nn\\&&
\chi(0,\mu)A_iT_i\chi(\lambda,0)A_j-
\chi(0,\mu)A_jT_j\chi(\lambda,0)A_i\,.
\eea
And then, multiplying this equation by the operator $A_kT_k$
from the left, taking sum by different permutations of
indices $i,j,k$ and using the first term of expansion
by $\lambda$, $\mu$ at $\lambda=0$, $\mu=0$, one obtains
the equation
\be
\sum_{(ijk)} \epsilon_{ijk}A_kT_k (\Delta_i u -uA_iT_iu)A_j=0,
\label{multiKP}
\ee
where $u=\chi_{\rm r}(0,0,{\bf x})$,
summation goes over different permutations of
indices.
This extremely
simple equation generates all equations of
multicomponent KP hierarchy (not containing
essentially discrete variables). Similar equation
can be found in \cite{Nijhoff}.

The scalar case of (\ref{multiKP}) represents a scalar
KP hierarchy in the form of superposition formula
for three B\"acklund transformations.
To simplify
the derivation of equations of KP hierarchy from this equation,
it is possible to take the first order of expansion
in terms of the parameter $a_1$, performing, in fact,
a transition $\Delta_1\rightarrow{\partial\over\partial x_1}$,
$T_1\rightarrow 1$ and
thus distinguishing
the continuous variable $x:=x_1$ (obtaining, in fact,
superposition formula for {\em two} B\"acklund
transformations).

It is easy to check that the choice
of three shifts
$T_i:{\bf x_i}\rightarrow {\bf x_i}+[a_i]$
in the formula (\ref{multiKP}) in the first
nontrivial order of expansion ($a_ia_ja_k$)
leads to the standard form of three waves equation.

Using cross-differences of equations (\ref{KPB}),
it is possible to obtain equations for the wave
functions (`modified equations')
\be
\sum_{(ijk)}\epsilon_{ijk}T_k\left((A_jT_i\widetilde\Psi)
\cdot \widetilde \Psi^{-1}\right)A_k=0.
\label{KPwave}
\ee
Equations of this type for the DS case and N-wave equations
case were derived in \cite{Kon3}, see also the lattice form
in \cite{Nijhoff}.

And finally, it is easy to check using (\ref{multiKP})
that $\Phi$ satisfies the equation
\be
(T_2\Delta_1 \Phi)\cdot(T_2\Delta_3 \Phi)^{-1}\cdot
(T_3\Delta_2 \Phi)\cdot(T_3\Delta_1 \Phi)^{-1}\cdot
(T_1\Delta_3 \Phi)\cdot(T_1\Delta_2 \Phi)^{-1}=1.
\label{KPsingman}
\ee
This symmetric equation represents a multicomponent
singularity manifold equation hierarchy.
In the scalar case it reads
\be
(T_2\Delta_1 \Phi)(T_3\Delta_2 \Phi)(T_1\Delta_3 \Phi)=
(T_2\Delta_3 \Phi)(T_3\Delta_1 \Phi)(T_1\Delta_2 \Phi).
\ee
We would like to remind here the main points of the concept
of Combescure symmetry transformation. The term
Combescure symmetry transformation originates from
geometry, where it is used for the transformation
of surface, keeping all tangent vectors parallel
\cite{Darboux}. For the integrable Darboux system
describing triply conjugate systems of surfaces
this transformation has a very simple interpretation
in the contexts of theory of integrable systems,
and in fact a general class of integrable equations
(`modified' equations) possesses this type of transformation
(though the geometrical interpretation in this case
may be missing). Our concept of the
Combescure transformation is inspired mainly
by integrability, not geometry (see \cite{Comb}).

Integrable equations
for the wave functions (i.e. modified equations and
singularity manifold type equations) possess the symmetry
transformations corresponding to the freedom of choice
of the wave function for the same CBA function.
In spectral terms this symmetry is described by the change
of weight functions $\rho(\lambda)$, $\eta(\mu)$.
In terms of equations themselves this symmetry
is represented using the Combescure invariants.
The solutions are connected by the Combescure
symmetry transformation, if Combescure invariants
for them are the same. The Combescure symmetry
group consists of two subgroups (`left' and `right'),
corresponding to the change of one weight function
($f(\lambda)$ or $\widetilde f(\mu)$, see notations
(\ref{Phi}); in our case  $f(\lambda)$ and $\widetilde f(\mu)$
are matrix-valued functions, multiplication by
$f(\lambda)$ is from the right,
by $\widetilde f(\mu)$ is from the left,
and here we will use the terms `left'
and `right' Combescure invariants according to the rules
of matrix multiplication).
Singularity manifold type equations possess
three types of invariants: left, right and
complete invariants. Modified equations possess
only one type of invariants (left or right) and are
invariant under opposite Combescure subgroup.
In this sense the basic hierarchy is invariant
under the action of all Combescure group.

We will
write down some of the Combescure
invariants of equations (\ref{KPwave}),
(\ref{KPsingman}). Combescure invariants
for the equation (\ref{KPwave}) read
\be
C_{ij}=(\Delta_i\widetilde \Psi A_j- \Delta_j\widetilde \Psi A_i)
(\widetilde\Psi)^{-1}.
\ee
Left Combescure invariants for the equation
(\ref{KPsingman}) are given by
\be
C'_{ij}=(\Delta_j \Phi)^{-1}(\Delta_i\Phi).
\ee
\section{Generalized DS hierarchy}
Now we will consider the 2-component case of multicomponent
KP
hierarchy in more detail.
In this section we derive the DS hierarchy
and the hierarchy of corresponding linear
problems,
the mDS hierarchy and linear problems,
and in the end we present the compact form
of the hierarchy of Combescure transformation
invariants for the mDS hierarchy. To perform
all the derivations in terms of hierarchy,
we use the formulae containing two continuous
times ($x_1:=x_{(1)1}$ and $x_2=:x_{(2)1}$
and discrete shifts $[a_1]$ and $[a_2]$,
the first acting on ${\bf x_{(1)}}$, the
second on ${\bf x_{(2)}}$. These shifts change
all the times of the hierarchy, and the expansion
in $a_1$, $a_2$ will generate all the hierarchy
in terms of partial differential equations.

To derive DS hierarchy and linear problems,
we will use the following special cases of the formula
(\ref{KPbasic21})
\bea
&&
{1\over a_1}\Delta_1\chi(\lambda,\mu,{\bf x})+
K_1(\lambda)\chi(\lambda,\mu,{\bf x})-
K_1(\mu)T_1\chi(\lambda,\mu,{\bf x})
\nn\\&&
=\widetilde\chi_1(\mu,{\bf x})\,T_1\chi_1(\lambda,{\bf x}),
\label{DSB1}
\\
&&
{1\over a_2}\Delta_2\chi(\lambda,\mu,{\bf x})+
K_2(\lambda)\chi(\lambda,\mu,{\bf x})-
K_2(\mu)T_2\chi(\lambda,\mu,{\bf x})
\nn\\&&
=
\widetilde\chi_2(\mu,{\bf x})\,T_2\chi_2(\lambda,{\bf x}),
\label{DSB2}
\\
&&
\partial_1\chi(\lambda,\mu,{\bf x})+
(K_1(\lambda)-
K_1(\mu))\chi(\lambda,\mu,{\bf x})=
\widetilde\chi_1(\mu,{\bf x})\,\chi_1(\lambda,{\bf x}),
\label{DSB3}
\\
&&
\partial_2\chi(\lambda,\mu,{\bf x})+
(K_2(\lambda)-
K_2(\mu))\chi(\lambda,\mu,{\bf x})=
\widetilde\chi_2(\mu,{\bf x})\,\chi_2(\lambda,{\bf x}),
\label{DSB4}
\eea
where we have introduced shift operators
and difference operators
\bea
&&
T_1\chi(\lambda,\mu,{\bf x_1},{\bf x_2})
=\chi(\lambda,\mu,{\bf x_1}+[a_1],{\bf x_2}),
\nn\\&&
T_2\chi(\lambda,\mu,{\bf x_1},{\bf x_2})
=\chi(\lambda,\mu,{\bf x_1},{\bf x_2}+[a_2]),
\nn\\&&
\Delta_1\chi(\lambda,\mu,{\bf x_1},{\bf x_2})
=(T_1-1)\chi(\lambda,\mu,{\bf x_1},{\bf x_2}),
\nn\\&&
\Delta_2\chi(\lambda,\mu,{\bf x_1},{\bf x_2})
=(T_2-1)\chi(\lambda,\mu,{\bf x_1},{\bf x_2}).
\nn
\eea
These operators can also be represented
as differential operators
$$
T_1=\exp\left(\sum_{n=1}^{\infty}
{a_1^n\over n}\partial_{(1)n}\right),
$$
$$
T_2=\exp\left(\sum_{n=1}^{\infty}
{a_2^n\over n}\partial_{(2)n}\right).
$$
The coefficient of expansion of $T_1$ in $a_1$,
$T_2$ in $a_2$ are given explicitly in terms of
Schur's polynomials
$p_N({\partial_{(1)1}\over 1},\dots,{\partial_{(1)N}\over N})$,
$p_N({\partial_{(2)1}\over 1},\dots,{\partial_{(2)N}\over N})$

The L-operator for the DS hierarchy is obtained by taking
(\ref{DSB3}) at $\lambda=0_2$, (\ref{DSB4}) at $\lambda=0_1$
(and then passing on to the wave functions (\ref{Phi}))
\bea
&&
\partial_1 \Psi_2=u\Psi_1,\quad u=\chi_{21}({\bf x});
\nn\\&&
\partial_2 \Psi_1=v\Psi_2,\quad v=\chi_{12}({\bf x});
\label{LDS}
\eea
while the hierarchy of A operators
is obtained by taking respectively
(\ref{DSB1}) at $\mu=0_2$, (\ref{DSB2}) at $\mu=0_1$
\bea
&&
{\Delta_1\over a_1}\Psi_2= uT_1\Psi_1\,,
\nn\\&&
{\Delta_2\over a_2}\Psi_1= vT_2\Psi_2\,.
\eea
One can also obtain dual L and A operators,
taking (\ref{DSB1}), (\ref{DSB3}) at $\mu=0_2$,
(\ref{DSB2}), (\ref{DSB4}) at $\lambda=0_1$.

It is possible to get DS hierarchy through
compatibility conditions, but we prefer to extract it
directly from the basic formulae (\ref{DSB1}),
(\ref{DSB2}), (\ref{DSB3}), (\ref{DSB4}).

Taking the difference of equations (\ref{DSB1}),
(\ref{DSB3}) at $\lambda=0_1$, $\mu=0_2$, and also
using similar combinations, one obtains
the system of equations
\bea
&&
({1/a_1}(T_1-1)-\partial_1)u=(T_1-1)\chi_{11}u,
\nn\\&&
({1/a_1}(1-T_1^{-1})-\partial_1)v=(T_1^{-1}-1)\chi_{11}v,
\nn\\&&
({1/a_2}(T_2-1)-\partial_2)v=(T_2-1)\chi_{22}v,
\nn\\&&
({1/a_2}(1-T_2^{-1})-\partial_2)u=(T_2^{-1}-1)\chi_{22}u,
\nn\\&&
\partial_1\chi_{22}=uv,
\nn\\&&
\partial_2\chi_{11}=uv.
\label{DShierarchy}
\eea
This system of equations represent a compact form
of DS hierarchy. The expansion of this systems
in $a_1$, $a_2$ generates the DS equation and higher
equations of the hierarchy. It is useful to note that
this expansion has the structure of recursion relations,
expressing the derivatives in higher times through
derivatives in lower times.

It is possible to obtain DS hierarchy from
the formula (\ref{multiKP}).
One could start with the  three
shifts $T_1:{\bf x_1}\rightarrow {\bf x_1}+[a_1]$,
$T_2:{\bf x_2}\rightarrow {\bf x_2}+[a_2]$,
$T_3:{\bf x_1}\rightarrow {\bf x_1}+[a_3]$
(or $T_3:{\bf x_2}\rightarrow {\bf x_2}+[a_3]$
for the part of hierarchy with higher times
connected with the disc $D_2$). Then, taking
the  first order of expansion in $a_1$,
$a_2$ and thus performing a transition
from two difference operators to derivatives
$\Delta_1\rightarrow{\partial\over\partial x_1}$,
$T_1\rightarrow 1$,
$\Delta_2\rightarrow{\partial\over\partial x_2}$,
$T_2\rightarrow 1$ and finally writing down matrix components
of equation (\ref{multiKP}), one obtains DS hierarchy
exactly in the form (\ref{DShierarchy}).

The zero order of expansion just gives trivial identities.
The first order of expansion of the first pair of equations in
$a_1$, the second pair in $a_2$ gives the DS
system
\bea
&&
\mbox{${1\over2}$}\left(\partial_{(1)2}+\partial_1^2\right)u=
u\partial_1 \chi_{11},
\nn\\&&
\mbox{${1\over2}$}\left(\partial_{(1)2}-\partial_1^2\right)v=
-v\partial_1 \chi_{11},
\nn\\&&
\mbox{${1\over2}$}\left(\partial_{(2)2}+\partial_2^2\right)v=
v\partial_2 \chi_{22},
\nn\\&&
\mbox{${1\over2}$}\left(\partial_{(2)2}-\partial_2^2\right)u=
-u\partial_2 \chi_{22},
\nn\\&&
\partial_1\chi_{22}=uv,
\nn\\&&
\partial_2\chi_{11}=uv,
\label{DSE}
\eea
where $\partial_{(1)2}={\partial\over\partial x_{(1)2}}$,
$\partial_{(2)2}={\partial\over\partial x_{(2)2}}$.
The usual form of the DS system (see e.g. \cite{solitons})
is a closed subsystem of (\ref{DSE}) written in terms of the
time $t$ defined as $\partial_t=\partial_{(1)2} +\partial_{(2)2}$.

The second order of expansion, with the use of
the DS equations (\ref{DSE}), gives
\bea
&&
\mbox{${1\over3}$}(\partial_{(1)3}-\partial_1^3)u=
u\partial_2^{-1}\partial_1(u\partial_1 v)-\partial_1(u
\partial_2^{-1}\partial_1 uv),
\nn\\&&
\mbox{${1\over3}$}(\partial_{(1)3}-\partial_1^3)v=
v\partial_2^{-1}\partial_1(v\partial_1 u)-\partial_1(v
\partial_2^{-1}\partial_1 uv),
\label{Veselov+}
\\&&
\mbox{${1\over3}$}(\partial_{(2)3}-\partial_2^3)v=
v\partial_1^{-1}\partial_2(v\partial_2 u)-\partial_2(v
\partial_1^{-1}\partial_2 uv),
\nn\\&&
\mbox{${1\over3}$}(\partial_{(2)3}-\partial_2^3)u=
u\partial_1^{-1}\partial_2(u\partial_2 v)-\partial_2(u
\partial_1^{-1}\partial_2 uv),
\label{Veselov-}
\eea
which is the higher DS system (\cite{mVN}).
The reduction $u=1$ gives rise to the split Nizhnik-Veselov-Novikov
equation for $v$ \cite{Nizhnik}, \cite{Veselov},
while the reduction $u=v$ leads to the modified
Nizhnik-Veselov-Novikov equation \cite{mVN}.

To proceed to modified DS hierarchy case, we integrate
equations (\ref{DSB1}),
(\ref{DSB2}), (\ref{DSB3}), (\ref{DSB4}) over $\lambda$,
$\mu$ with some weight functions and obtain
equations for wave functions
\bea
&&
{\Delta_1\over a_1}
\Phi=\widetilde \Psi_1 T_1 \Psi_1,
\nn\\&&
{\Delta_2\over a_2}
\Phi=\widetilde \Psi_2 T_2 \Psi_2,
\nn\\&&
\partial_1\Phi=\widetilde \Psi_1 \Psi_1,
\nn\\&&
\partial_2\Phi=\widetilde\Psi_2\Psi_2.
\label{DSwave}
\eea
The L operator for modified DS hierarchy can be obtained
from the last pair of equations, using equations
(\ref{LDS}) and dual DS L-operator
\be
\partial_1\partial_2\Phi=(\partial_2 U)\partial_1 \Phi+
(\partial_1 V)\partial_2 \Phi,
\label{LMDS}
\ee
where $U=\ln \Psi_1$, $V=\ln \Psi_2$.
The hierarchy of A operators arises if
one uses the second pair of equations (\ref{DSwave})
to exclude $\widetilde \Psi_1$,
$\widetilde \Psi_2$ from the first pair
\bea
&&
{\Delta_1\over a_1}\Phi=W_1\partial_1\Phi,\quad
W_1={T_1\Psi_1\over\Psi_1};
\nn\\&&
{\Delta_2\over a_2}\Phi=W_2\partial_2\Phi,\quad
W_2={T_2\Psi_2\over\Psi_2}.
\label{AMDS}
\eea
The compatibility condition for (\ref{LMDS}),
(\ref{AMDS}) gives the modified DS hierarchy
\bea
&&
{1\over a_1}\Delta_1 \partial_2 U+(T_1\partial_2 U)(\partial_1 W_1)
+(T_1\partial_1 V)(\partial_2 W_1)+
W_1(\partial_2 U)\Delta_1\partial_1 V=
\nn\\&&
\partial_1\partial_2 W_1+
\partial_1(W_1\partial_2 U),
\nn\\&&
{1\over a_1}\Delta_1 \partial_1 V + W_1(\Delta_1\partial_1 V)
(\partial_1 V)-\partial_1(W_1\partial_1 V)=0,
\nn\\&&
W_1=\exp(\Delta_1 U),
\eea
and for times with the subscript $(2)$
\bea
&&
{1\over a_2}\Delta_2 \partial_1 V+(T_2\partial_1 V)(\partial_2 W_2)
+(T_2\partial_2 U)(\partial_1 W_2)+
W_2(\partial_1 V)\Delta_2\partial_2 U=
\nn\\&&
\partial_2\partial_1 W_2+
\partial_2(W_2\partial_1 V),
\nn\\&&
{1\over a_2}\Delta_2 \partial_2 U + W_2(\Delta_2\partial_2 U)
(\partial_2 U)-\partial_2(W_2\partial_2 U)=0,
\nn\\&&
W_2=\exp(\Delta_2 V).
\eea
The zero order of expansion of these equations
in $a_1$, $a_2$
gives
modified DS equation \cite{Kon3}
\bea
&&
\mbox{${1\over2}$}\partial_2(\partial_{(1)2}U-\partial_1^2U
-(\partial_1 U)^2)
+\partial_1(\partial_2 U\cdot\partial_1V)=0,
\nn\\&&
\partial_{(1)2}V+\partial_1^2V+(\partial_1V)
=\partial_1V\cdot\partial_1U,
\\&&
\mbox{${1\over2}$}\partial_1(\partial_{(2)2}V-\partial_2^2V-
(\partial_2 V)^2)
+\partial_2(\partial_2 U\cdot\partial_1 V)=0,
\nn\\&&
\partial_{(2)2}U+\partial_2^2U+(\partial_2U)^2
=\partial_2V\cdot\partial_2U.
\eea
The hierarchy of Combescure invariants (\cite{Comb})
for the modified DS hierarchy
is obtained by representation of
the DS solutions $u,v$ through the mDS solutions
$U,V$
\bea
&&
u=\exp(-T_1U){\Delta_1\over a_1}\exp(V),
\nn\\&&
v=\exp(-T_2V){\Delta_2\over a_2}\exp(U).
\eea
\section{$\tau$-function for the generalized DS hierarchy}
Now we will analyze the functional equations (\ref{KPbasic0})
for the case $N=2$. For convenience we present them
here explicitly:
\begin{eqnarray}
&&\left({\frac{\mu-a_{1}}{\mu-b_{1}}}\right)
\chi_{11}(\lambda,\mu,{\bf x}+[{\bf a}])- \left({
\frac{\lambda-a_{(1)}}{\lambda-b_{(1)}}}\right)
\chi_{11}(\lambda,\mu,{\bf x}+[{\bf b}])+  \nonumber
\\
&&
(b_{(1)}-a_{(1)})
\chi_{1\,1}(\lambda,b_{(1)},{\bf x}+[{\bf a}])
\chi_{1\,1}(b_{(1)},\mu,{\bf x}+[{\bf b}])+\nn\\
&&(b_{(2)}-a_{(2)})
\chi_{2\,1}(\lambda,b_{(2)},{\bf x}+[{\bf a}])
\chi_{1\,2}(b_{(2)},\mu,{\bf x}+[{\bf b}])=0,
\label{DSbasic01}
\\
&&\left({\frac{\mu-a_{(2)}}{\mu-b_{(2)}}}\right)
\chi_{22}(\lambda,\mu,{\bf x}+[{\bf a}])- \left({
\frac{\lambda-a_{(2)}}{\lambda-b_{(2)}}}\right)
\chi_{22}(\lambda,\mu,{\bf x}+[{\bf b}])+  \nonumber
\\
&&
(b_{(1)}-a_{(1)})
\chi_{1\,2}(\lambda,b_{(1)},{\bf x}+[{\bf a}])
\chi_{2\,1}(b_{(1)},\mu,{\bf x}+[{\bf b}])+\nn\\
&&(b_{(2)}-a_{(2)})
\chi_{2\,2}(\lambda,b_{(2)},{\bf x}+[{\bf a}])
\chi_{2\,2}(b_{(2)},\mu,{\bf x}+[{\bf b}])=0,
\label{DSbasic02}
\\
&&\left({\frac{\mu-a_{(1)}}{\mu-b_{(1)}}}\right)
\chi_{12}(\lambda,\mu,{\bf x}+[{\bf a}])- \left({
\frac{\lambda-a_{(2)}}{\lambda-b_{(2)}}}\right)
\chi_{12}(\lambda,\mu,{\bf x}+[{\bf b}])+  \nonumber
\\
&& (b_{(1)}-a_{(1)})
\chi_{1\,2}(\lambda,b_{(1)},{\bf x}+[{\bf a}])
\chi_{1\,1}(b_{(1)},\mu,{\bf x}+[{\bf b}])+\nn\\
&&(b_{(2)}-a_{(2)})
\chi_{2\,2}(\lambda,b_{(2)},{\bf x}+[{\bf a}])
\chi_{1\,2}(b_{(2)},\mu,{\bf x}+[{\bf b}])=0,
\label{DSbasic03}
\\
&&\left({\frac{\mu-a_{(2)}}{\mu-b_{(2)}}}\right)
\chi_{21}(\lambda,\mu,{\bf x}+[{\bf a}])- \left({
\frac{\lambda-a_{(1)}}{\lambda-b_{(1)}}}\right)
\chi_{21}(\lambda,\mu,{\bf x}+[{\bf b}])+  \nonumber
\\
&&
(b_{(1)}-a_{(1)})
\chi_{1\,1}(\lambda,b_{(1)},{\bf x}+[{\bf a}])
\chi_{2\,1}(b_{(1)},\mu,{\bf x}+[{\bf b}])+\nn\\
&&(b_{(2)}-a_{(2)})
\chi_{2\,1}(\lambda,b_{(2)},{\bf x}+[{\bf a}])
\chi_{2\,2}(b_{(2)},\mu,{\bf x}+[{\bf b}])=0.
\label{DSbasic04}
\end{eqnarray}
First it is clear that equations (\ref{DSbasic01})
with $a_2=b_2$ and (\ref{DSbasic02})
with $a_1=b_1$ which coincide with those for scalar KP
hierarchy give rise to the following expressions for
$\chi_{11}$ and $\chi_{22}$ (see \cite{KP1})
\be
\chi_{11}(\lambda,\mu,{\bf x})={1\over(\lambda-\mu)}
{\tau_1({\bf x_1}-[\lambda]+[\mu],{\bf x_2})
\over \tau_1({\bf x})},
\label{tauform}
\ee
\be
\chi_{22}(\lambda,\mu,{\bf x})={1\over(\lambda-\mu)}
{\tau_2({\bf x_1},{\bf x_2}-[\lambda]+[\mu])
\over \tau_2({\bf x})}
\label{tauform'},
\ee
where $\tau_1$ and $\tau_2$ are the components
of the $\tau$-function which depend on the variables
${\bf x}=({\bf x_1},{\bf x_2})$.

Let us consider now equation (\ref{DSbasic02}) at
$\mu=a_1$ and $a_2=b_2$. Substituting the expression
(\ref{tauform}) into this equation, we obtain
\bea
\tau_1({\bf x_1}+[b_1], {\bf x_2}+[a_2])
\chi_{12}(\lambda,a_1,{\bf x_1}+[b_1], {\bf x_2}+[a_2])=\nn\\
\tau_1({\bf x_1}+[a_1], {\bf x_2}+[a_2])
\chi_{12}(\lambda,a_1,{\bf x_1}+[a_1], {\bf x_2}+[a_2]).
\label{DStau1}
\eea
For $b_1=0$ after the change
${\bf x_2}+[a_2]\rightarrow {\bf x_2}$, one gets
\be
\tau_1({\bf x_1}, {\bf x_2})
\chi_{12}(\lambda,a_1,{\bf x_1}, {\bf x_2})=
\tau_1({\bf x_1}+[a_1], {\bf x_2})
\chi_{12}(\lambda,0,{\bf x_1}+[a_1], {\bf x_2}).
\ee
Denoting the r.h.s. of this equation as
$\tilde\tau_{12}(\lambda,{\bf x_1}+[a_1], {\bf x_2})$,
we rewrite it as
\be
\chi_{12}(\lambda,\mu,{\bf x_1}, {\bf x_2})=
{\tilde\tau_{12}(\lambda,{\bf x_1}+[\mu], {\bf x_2})\over
\tau_1({\bf x_1}, {\bf x_2})}.
\label{DStau2}
\ee
Then the substitution of (\ref{DStau2}) into equation
(\ref{DStau1}) with $\mu=a_1$ and $\lambda=a_2$ gives rise to
\bea
\tau_1({\bf x_1}+[a_1], {\bf x_2}+[b_2])
{\tilde\tau_{12}(a_2,{\bf x_1}+[a_1]+[b_1], {\bf x_2}+[a_2])
\over
\tau_1({\bf x_1}+[a_1], {\bf x_2}+[a_2])}=\\
\tau_2({\bf x_1}+[a_1], {\bf x_2}+[b_2])
{\tilde\tau_{12}(b_2,{\bf x_1}+[a_1]+[b_1], {\bf x_2}+[b_2])
\over
\tau_1({\bf x_1}+[a_1], {\bf x_2}+[a_2])}.
\eea
This equation implies that
\be
\tilde\tau_{12}(a,{\bf x_1},{\bf x_2})=
{\tau_{12}({\bf x_1},{\bf x_2}-[a])
\over
\rho({\bf x_1},{\bf x_2})}
\label{DStau3}
\ee
and
\be
\rho({\bf x_1},{\bf x_2})=
{\tau_2({\bf x_1},{\bf x_2})
\over
\tau_1({\bf x_1},{\bf x_2})}.
\ee
where $\tau_{12}$ is an arbitrary function.
Substitution of (\ref{DStau3}) into (\ref{DStau2}) gives
\be
\chi_{12}(\lambda,\mu,{\bf x_1}, {\bf x_2})=
{\tau_{12}(\lambda,{\bf x_1}+[\mu], {\bf x_2}-[\lambda])\over
\tau_2({\bf x_1}, {\bf x_2})}.
\label{tau2}
\ee
Now we proceed with equation  (\ref{DSbasic04}).
Substitution of the expression  (\ref{tauform'}) into this equation
with $\mu=a_2$, $a_1=b_1$ gives
\bea
&&
\chi_{21}(\lambda,a_2,{\bf x_1}+[a_1], {\bf x_2}+[b_2])=
\nn\\&&
{\tau_2({\bf x_1}+[a_1], {\bf x_2}+[a_2])
\over
\tau_2({\bf x_1}+[a_1], {\bf x_2}+[b_2])}
\chi_{21}(\lambda,b_2,{\bf x_1}+[a_1], {\bf x_2}+[a_2])
\eea
After the change ${\bf x_1}+[a_1]\rightarrow {\bf x_1} $ and the
choice $b_2=0$ one gets
\be
\tau_2({\bf x})\chi_{21}(\lambda,a_2,{\bf x})=
\tau_2({\bf x_1}, {\bf x_2}+[a_2])\chi_{21}
(\lambda,0,{\bf x_1}, {\bf x_2}+[a_2]).
\ee
From this equation one concludes that
\be
\chi_{21}(\lambda,\mu,{\bf x})=
{\tilde\tau_{21}(\lambda,{\bf x_1}, {\bf x_2}+[\mu])
\over
\tau_2({\bf x})},
\label{DStau4}
\ee
where $\tilde\tau_{21}(\lambda,{\bf x_1}, {\bf x_2})=
\tau_2({\bf x})\chi_{21}(\lambda,0,{\bf x})$.

Substituting now
with $\mu=a_2$, $\lambda=a_1$, $b_2=0$
we obtain
\be
{\tau_1({\bf x_1}+[a_1],{\bf x_2})
\over
\tau_2({\bf x_1}+[a_1],{\bf x_2})}
\tilde\tau_{21}(a_1,{\bf x_1}+[a_1],{\bf x_2})=
{\tau_1({\bf x_1}+[b_1],{\bf x_2})
\over
\tau_2({\bf x_1}+[b_1],{\bf x_2})}
\tilde\tau_{21}(b_1,{\bf x_1}+[b_1],{\bf x_2}).
\label{DStau01}
\ee
Hence
\be
\tilde\tau_{21}(a_1,{\bf x_1},{\bf x_2})=
{\tau_{21}({\bf x_1}-[a_1],{\bf x_2})
\over
\tilde\rho({\bf x})}
\label{DStau5}
\ee
and
\be
\tilde\rho({\bf x})=
{\tau_1({\bf x})
\over
\tau_2({\bf x})}.
\label{DStau6}
\ee
Substitution of (\ref{DStau5}) and (\ref{DStau6})
into (\ref{DStau4}) gives
\be
\chi_{21}(\lambda,\mu,{\bf x})=
{\tau_{21}({\bf x_1}-[\lambda],{\bf x_2}+[\mu])
\over
\tau_1({\bf x})}.
\ee
To establish the relation between $\tau_1$ and $\tau_2$
we consider again equation (\ref{DStau01}).
Due to (\ref{DStau5}) one gets
\be
{\tau_1({\bf x_1}+[b_1],{\bf x_2})
\over
\tau_2({\bf x_1}+[b_1],{\bf x_2})}=
{\tau_1({\bf x_1}+[a_1],{\bf x_2})
\over
\tau_2({\bf x_1}+[a_1],{\bf x_2})}.
\ee
Since $a_1$ and $b_1$ are arbitrary parameters, one concludes that
\be
{\tau_1({\bf x})
\over
\tau_2({\bf x})}=
f({\bf x_2}).
\ee
In the same way, it follows from (\ref{DSbasic03}) that
\be
{\tau_1({\bf x})
\over
\tau_2({\bf x})}=
f({\bf x_1}).
\ee
Thus
\be
{\tau_1({\bf x})
\over
\tau_2({\bf x})}=
{\rm const}.
\ee
Having in mind the formulae (\ref{tauform}) and (\ref{tauform'}),
one can put $\tau_1=\tau_2$ without the loss of generality.

So for the 2-component KP hierarchy one has
\be
\chi_{11}(\lambda,\mu,{\bf x})={1\over(\lambda-\mu)}
{\tau({\bf x_1}-[\lambda]+[\mu],{\bf x_2})
\over \tau({\bf x})},
\label{tauform1}
\ee
\be
\chi_{22}(\lambda,\mu,{\bf x})={1\over(\lambda-\mu)}
{\tau({\bf x_1},{\bf x_2}-[\lambda]+[\mu])
\over \tau({\bf x})},
\label{tauform2}
\ee
\be
\chi_{12}(\lambda,\mu,{\bf x})=
{\tau_{12}({\bf x_1}+[\mu],{\bf x_2}-[\lambda])
\over
\tau({\bf x})},
\label{tauform3}
\ee
\be
\chi_{21}(\lambda,\mu,{\bf x})=
{\tau_{21}({\bf x_1}-[\lambda],{\bf x_2}+[\mu])
\over
\tau({\bf x})}.
\label{tauform4}
\ee
At this step $\tau$, $\tau_{12}$ and $\tau_{21}$
are arbitrary functions. Substitution of the expressions
(\ref{tauform1}-\ref{tauform4})
into the general functional equations
(\ref{DSbasic01}-\ref{DSbasic04})
gives rise to the four bilinear functional
equations for these functions which represent
themselves the addition formulae. We will write down them
explicitly for the multicomponent case later.
\section{$\tau$-function and addition formulae for
n-component KP hierarchy}
It will be shown in the next section that
the objects of multicomponent KP hierarchy depend
not only on continuous times, but also on
discrete times (indices). The addition formulae
arising from the discrete dynamics will be considered
in the context of Toda lattice hierarchy. Here we
derive only the addition formulae following
from the basic formula (\ref{KPbasic0})
defining the dynamics for continuous variables.

For the general multicomponent KP hierarchy the system
(\ref{KPbasic0}) looks rather complicated. But its
analysis is simplified considerably due to the
special structure of equations (\ref{KPbasic0}).
It isn't difficult to see that for two fixed values
of $\alpha$ and $\beta$ , taking $a_{\gamma}=b_{\gamma}$,
$\gamma\neq \alpha$, $\gamma\neq\beta$, the system reduces to
four equations of the DS hierarchy (\ref{DSbasic01}-\ref{DSbasic04})
for the variables ${\bf x_{\alpha}}$ and  ${\bf x_{\beta}}$.
Hence, in these variables we have the formulae
\bea
&&
\chi_{\alpha\alpha}(\lambda,\mu,{\bf x})={1\over(\lambda-\mu)}
{\tau({\bf x_1},...,{\bf x_{\alpha}}-[\lambda]+[\mu],...,{\bf x_N})
\over \tau({\bf x})},
\nn\\
&&
\chi_{\alpha\beta}(\lambda,\mu,{\bf x})=
{\tau_{\alpha\beta}({\bf x_1},...,{\bf x_\alpha}+[\mu],...,
{\bf x_\beta}-[\lambda],...,{\bf x_N})
\over
\tau({\bf x})}\;\;\;(\alpha\neq\beta).
\label{tauformN}
\eea
Here indices $\alpha$ and $\beta$ are arbitrary. Therefore the
formulae  (\ref{tauformN}) take place for all $\alpha$ and $\beta$.

To compactify these formulae we will introduce the shift
operator $T_{\alpha}(\lambda)$ ($\alpha=1,...,N$)
\be
T_{\alpha}(\lambda)f({\bf x})=
f({\bf x_1},...,{\bf x_\alpha}+[\lambda],...,
{\bf x_N}).
\ee
Using such operators, one can present the formulae
(\ref{tauformN}) in the form
\bea
\chi_{\alpha\alpha}(\lambda,\mu,{\bf x})=
{1\over\lambda-\mu}
{T_{\alpha}(\mu)T_{\alpha}^{-1}(\lambda)\tau({\bf x})
\over \tau({\bf x})}
\quad \alpha=1,...,N\,,
\nn\\
\chi_{\alpha\beta}(\lambda,\mu,{\bf x})=
{1\over\lambda-\mu}
{T_{\alpha}(\mu)T_{\beta}^{-1}(\lambda)\tau_{\alpha\beta}({\bf x})
\over \tau({\bf x})}
\quad \alpha\neq\beta\,.
\eea
In the special cases $\lambda=0$ or $\mu=0$
these formulae reduce to the standard formulae for
the Baker-Akhiezer function \cite{Date}, \cite{Orlov},
\cite{Dickey}.
Separating the diagonal case $\alpha=\beta$
and offdiagonal one, one gets
\bea
&&(a_\alpha-\mu)(\lambda-b_\alpha)
\tau({\bf x}+[{\bf b}])
T_\alpha(\mu)
T_\alpha^{-1}(\lambda)
\tau({\bf x}+[{\bf a}])+
\nn\\
&&(b_\alpha-\mu)(\lambda-a_\alpha)
\tau({\bf x}+[{\bf a}])
T_\alpha(\mu)
T_\alpha^{-1}(\lambda)
\tau({\bf x}+[{\bf b}])+
\nn\\
&&(b_\alpha-a_\alpha)(\lambda-\mu)
T_\alpha(\mu)
T_\alpha^{-1}(b_\alpha)
\tau({\bf x}+[{\bf b}])
T_\alpha(b_\alpha)
T_\alpha^{-1}(\lambda)
\tau({\bf x}+[{\bf a}])+
\nn\\
&&(\mu-\lambda)(\mu-b_\alpha)(\lambda-b_\alpha)
\sum_{\gamma\neq\alpha}
(b_\gamma-a_\gamma)
T_\alpha(\mu)
T_\gamma^{-1}(b_\gamma)
\tau_{\alpha\gamma}({\bf x}+[{\bf b}])\times
\nn\\
&&T_\gamma(b_\gamma)
T_\beta^{-1}(\lambda)
\tau_{\gamma\beta}({\bf x}+[{\bf a}])=0
\label{addition1}
\eea
and
\bea
&&(\mu-a_\alpha)(\lambda-b_\beta)
\tau({\bf x}+[{\bf b}])
T_\alpha(\mu)
T_\beta^{-1}(\lambda)
\tau_{\alpha\beta}({\bf x}+[{\bf a}])+
\nn\\
&&(\mu-b_\alpha)(\lambda-a_\beta)
\tau({\bf x}+[{\bf a}])
T_\alpha(\mu)
T_\beta^{-1}(\lambda)
\tau_{\alpha\beta}({\bf x}+[{\bf b}])+
\nn\\
&&(b_\alpha-a_\alpha)(\lambda-b_\beta)
T_\alpha(\mu)
T_\alpha^{-1}(b_\alpha)
\tau({\bf x}+[{\bf b}])
T_\alpha(b_\alpha)
T_\beta^{-1}(\lambda)
\tau_{\alpha\beta}({\bf x}+[{\bf a}])+
\nn\\
&&(\mu-b_\alpha)(\lambda-b_\beta)
\sum_{\gamma\neq\alpha,\beta}
(b_\gamma-a_\gamma)
T_\alpha(\mu)
T_\gamma^{-1}(b_\gamma)
\tau_{\alpha\gamma}({\bf x}+[{\bf b}])\times
\nn\\
&&T_\gamma(b_\gamma)
T_\beta^{-1}(\lambda)
\tau_{\gamma\beta}({\bf x}+[{\bf a}])=0.
\label{addition2}
\eea
Equations (\ref{addition1}) and  (\ref{addition2})
represent the extension of the simplest addition formula
for the $\tau$-function in the scalar case (see \cite{KP1})
to the multicomponent case. They are related
to the generalization of the bilinear relations for the
Pl\"ucker coordinates to the multicomponent case. On the
other hand they could be connected with the possible
multidimensional extensions of the Fay's trisecant
formula \cite{Fay}.

More general addition formulae for the $\tau$-function
follow from the general functional equation for
the
function $\chi_{\alpha\beta}$ which corresponds to the
choice
\be
(g_1({\bf x}) g_2^{-1}({\bf x}))_{\alpha\beta}
=\delta_{\alpha\beta}
\prod_{k=1}^{N}
{\nu-a_{\alpha k}
\over
\nu-b_{\alpha k}}.
\ee
\section{$\tau$-functional substitution
and closed 1-form}
The formulae (\ref{tauform1}), (\ref{tauform2}), (\ref{tauform3}),
(\ref{tauform4}) indicate that it is possible to preserve
in multicomponent case
the
general formula connecting the $\tau$-function and the CBA
function in the scalar case. It would be more consistent
first to define the $\tau$-function through the CBA function
in terms of the closed one-form, and then to prove that
this definition implies the general formula.
On the other hand, one may treat the general formula
as the definition and then prove the correctness
of this definition reducing it to closed 1-form
and showing the equivalence of `local'
and `global' formulae
(that implies unique existence of the $\tau$-function,
at least in some neighborhood
of $g=1$) .
In this paper we would rather
follow the second way.

So we define the $\tau$-function by the relation
\be
\chi(\lambda,\mu,g)=\chi_0(\lambda,\mu){\tau(g_{(\lambda,\mu)}g)
\over
\tau(g)}.
\label{tau-substitution}
\ee
Here  $\chi_0(\lambda,\mu)$ is just the function
$(\lambda-\mu)^{-1}$ extended as 1 for $\lambda$, $\mu$
in different $D_i$, and the function $g_{(\lambda,\mu)}$
is an extension of the function ${\nu-\lambda\over\nu-\mu}$
to the case of disconnected $G$
\bea
&&\chi_0(\lambda,\mu)=
\left\{
\begin{array}{cl}
{1\over\lambda-\mu} &\mbox{ for }\lambda\in D_i,\mu\in D_i\\
1&\mbox{ for }\lambda\in D_i,\mu\in D_j,i\neq j\,.
\end{array}
\right.
\eea
The function  $g_{(\lambda,\mu)}$ (
in fact, it is a set of $N$ loop group elements
$g_i(\nu)$ depending on the parameters $\lambda,\mu$)
is the ratio of two terms
\bea
&&g_{(\lambda,\mu)}={g_{(\lambda)}\over g_{(\mu)}},
\nn\\
&&g_{(\lambda)\;i}(\nu)=
\left\{
\begin{array}{cl}
\nu -\lambda&\mbox{ for }\lambda\in D_i\\
1&\mbox{ for }\lambda\notin D_i\,.
\end{array}
\right.
\eea
To extract the formulae (\ref{tauform1}), (\ref{tauform2}), (\ref{tauform3}),
(\ref{tauform4}) from the formula (\ref{tau-substitution}),
one should recall that the loop group defining the
multicomponent KP hierarchy can be decomposed in the
following way
\be
g_i(\nu)=g_{i\;{\rm out}}(\nu)\times
\nu^{n_i},\quad \sum_{i=1}^{N}n_i=0,
\ee
where the first term defines the dependence on continuous
times and the second -- on the discrete times (indices).
If $\lambda$, $\mu$ belong to the same $D_i$, using
explicit parametrization of the loop group elements,
we obtain exactly the formulae  (\ref{tauform1}), (\ref{tauform2}).
For $\lambda\in D_k$, $\mu\in D_p$, $k\neq p$,
$g_{(\lambda,\mu)}$ has a factorization
\bea
&&
g_{(\lambda,\mu)\;k}(\nu)={\nu-\lambda\over\nu}\times\nu,
\nn\\
&&
g_{(\lambda,\mu)\;p}(\nu)={\nu\over\nu-\mu}\times\nu^{-1},
\nn\\
&&
g_{(\lambda,\mu)\;i}(\nu)=1,\quad i\neq k,i\neq p\,.
\eea
Using these formulae, we obtain the formulae
(\ref{tauform3}), (\ref{tauform4}), with the components
of $\tau$-function interpreted as a $\tau$-function
with shifted index variables (general index zero is
preserved). For example,
in the two-component case the function $\tau_{12}$
corresponds to $\tau(n_1-1,n_2+1)$. To express
the CBA function with zero indices, we need to know
the set of functions $\tau$, $\tau_{ij}$, $i\neq j$.
Then for other admissible index vectors the CBA
function and the $\tau$-function can be found
explicitly through the determinant formulae
\cite{NLS}. Thus, the discrete dynamics is exactly solvable
in terms of initial data (CBA function with zero indices
or the set of $\tau$-functions  $\tau$, $\tau_{ij}$, $i\neq j$).

Returning to our original purpose, we will extract
the definition of the $\tau$-function in terms of the
closed 1-form from (\ref{substitution}) and then
prove that both definitions are equivalent.
Our derivation will mostly repeat the derivation
in the scalar case given in \cite{KP1},
but we would like to remind it due to the importance for
the following consideration.

Differentiating (\ref{tau-substitution}) with respect to $\lambda$,
one obtains
\be
-{1\over\tau(g)}\oint {\delta \tau\left(g\times
g_{(\lambda,\mu)}\right)
\over\delta \ln g_{\alpha}(\nu)}{1\over\nu-\lambda}d\nu
={\partial\over \partial \lambda}(\lambda-\mu)
\chi_{\alpha\alpha}(\lambda,\mu;g)\,,
\ee
or for $\lambda=\mu$
\be
-{1\over\tau(g)}\oint {\delta \tau(g)
\over\delta \ln g_\alpha(\nu)}{1\over\nu-\lambda}d\nu
=
\chi_{{\rm r}\,\alpha\alpha}(\lambda,\lambda;g),
\ee
where $\chi_{{\rm r}\,\alpha\beta}(\lambda,\mu;g)=
\chi_{\alpha\beta}(\lambda,\mu;g)- \delta_{\alpha\beta}
(\lambda-\mu)^{-1}$,`r' is for `regularized'.
This equation can be rewritten in the form
\be
-{1\over\tau(g)}\oint {\delta \tau(g)
\over\delta \ln g_\alpha(\nu)}{1\over\nu-\lambda}d\nu
=\oint
\chi_{{\rm r}\,\alpha\alpha}(\nu,\nu;g){1\over\nu-\lambda}d\nu\,.
\label{formvar1}
\ee
Relation (\ref{formvar1}) implies that the linear functionals
$-{1\over\tau(g)}{\delta \tau(g(\nu))\over\delta \ln g}
$ and
$\chi_{\rm r}(\nu,\nu;g)$
are identical for arbitrary variations  ${\delta g\over g}$
analytic outside each $D_i$ and
decreasing at infinity.
Thus for ${\delta g\over g}$ belonging to this class
of functions
the variation of the $\tau$-function
is given by
\be
-\delta \ln \tau(g)=
\sum_{\alpha=1}^N\oint
\chi_{{\rm r}\,\alpha\alpha}(\nu,\nu;g){\delta g_\alpha(\nu)\over
g_\alpha(\nu)} d\nu\,.
\label{formvar}
\ee
This expression defines a variational 1-form defining the
$\tau$-function. It is easy to prove using the identity (\ref{HIROTA})
that this form is
{\em closed} .
Indeed, according to (\ref{HIROTA}) (here we are using scalar
notations, it is not difficult to extract equations
in terms of components $\chi_{\alpha\beta}(\lambda,\mu;g)$)
\bea
\delta \chi(\lambda,\mu;g)=
\int_{\partial G} \chi(\nu,\mu;g){\delta g(\nu)\over g(\nu)}
\chi(\lambda,\nu;g)d\nu\,,\nn\\
\delta \chi_{\rm r}(\lambda,\lambda;g)=
\int_{\partial G}\chi_r(\nu,\lambda;g){\delta g(\nu)\over g(\nu)}
\chi_{\rm r}(\lambda,\nu;g)d\nu\,.
\label{variation1}
\eea
So the variation
of the (\ref{formvar}) gives
\bea
-\delta^2 \ln \tau(g)=
\int_{\partial G}\!\!\!\int_{\partial G}
\chi_{\rm r}(\nu,\lambda;g)\chi_{\rm r}
(\lambda,\nu;g){\delta' g(\nu)\over g(\nu)}
{\delta g(\lambda)\over g(\lambda)} d\nu\,d\lambda\,.
\eea
The symmetry of the kernel of second variation with respect to
$\lambda$, $\nu$ implies that the form (\ref{formvar})
is closed.

So the formula (\ref{formvar}) gives the definition of the
$\tau$-function in terms of the closed 1-form. For the standard KP
coordinates
$$
{\delta g_{\alpha}(\lambda)\over g_{\alpha}(\lambda)}=
\sum_{n=1}^{\infty} {dx_{(\alpha) n} \over \lambda^n}\,.
$$
This formula allows us to obtain a closed 1-form in terms
of $dx_{\alpha n}$
\be
-d \ln \tau({\bf k},{\bf x})=
\sum_{\alpha=1}^N\sum_{n=0}^{\infty}
\left.{\partial^n\over\partial\nu^n}
\chi_{{\rm r}\; \alpha\alpha}(\nu,\nu, {\bf k},{\bf x})
\right|_{\nu=0}dx_{(\alpha) n}\,.
\label{formvar4}
\ee

To prove that the `local' definition in terms of the closed
1-form implies the global definition (\ref{tau-substitution}),
it is sufficient to show that logarithmic derivatives
of the l.h.s. and the r.h.s.
of (\ref{tau-substitution}) with respect to $\lambda$, $\mu$
and variational logarithmic derivatives with respect to
$\delta g_\alpha/g_\alpha$ are equal in virtue
of (\ref{formvar}) and generalized Hirota bilinear
identity (\ref{HIROTA}).
Then the l.h.s. and the r.h.s.
of (\ref{tau-substitution})
could differ only by the factor, and the normalization
of the function $\chi$ implies that this factor is equal to 1.

Here we will prove only the equality of logarithmic
variations, the proof for derivatives is analogous
(see \cite{KP1}).

The logarithmic variation of (\ref{tau-substitution})
gives
\bea
&&\delta\left(\ln
\chi(\lambda,\mu;g)(\lambda-\mu)-\ln\tau(g\times
g_{(\lambda,\mu)})
+\ln\tau(g)\right)=
\nn\\
&&{1\over\chi(\lambda,\mu)}
\int_{\partial G} \chi(\nu,\mu;g){\delta g(\nu)\over g(\nu)}
\chi(\lambda,\nu;g)d\nu
\nn\\&&
-\int_{\partial G}
(\chi_{\rm r}(\nu,\nu;g\times g_{(\lambda,\mu)})-
\chi_{\rm r}(\nu,\nu;g))
{\delta g(\nu)\over g(\nu)} d\nu\nn\\
\eea
(we have used (\ref{variation1}) and (\ref{formvar})).
After the use of the determinant formula \cite{NLS}
\be
\chi(\lambda ,\mu;g\times g_{(\lambda',\mu')})
=g_{(\lambda',\mu')}^{-1}(\lambda) g_{(\lambda',\mu')}(\mu)
\frac{
\left|
\begin{array}{cc}
\chi (\lambda',\mu') & \chi (\lambda',\mu ) \\
\chi (\lambda ,\mu') & \chi (\lambda ,\mu )
\end{array}
\right|
}{\chi (\lambda',\mu')}
\label{determinant1}
\ee
to transform the second term, one concludes
that the variation is equal to zero.

\section{Generalized 2D Toda lattice hierarchy. Basic equations.}
We will consider here for simplicity only scalar
2D Toda lattice hierarchy. It is generated by the
generalized Hirota identity with the domain $G$
which consists of two disconnected regions, i.e. in
some sense it is a two-point KP hierarchy.
But it also possess some fundamental special
features and can be treated as a primary
object with respect to scalar KP hierarchy.

Without loss of generality one can choose
$G$ as
$$
G=D_0\cup D_\infty\,,
$$
where $D_0$ and $D_\infty$ are unit discs around
the origin and infinity respectively. The
two-component KP hierarchy is generated by the
pair of functions  $g_{\rm in}$, $g_{\rm out}$ defined
on the boundary of $D_0$, $D_\infty$ respectively,
having zero general index (let their individual indices
be $n$, $-n$),  $g_{\rm out}$  is analytic and have no zeroes outside
the unit circle, $g_{\rm in}$ inside the unit circle.
It is easy to see, that dynamics defined by the pair
of these functions coincides with the dynamics defined
by one function for both boundaries
\be
g=\lambda^n g_{\rm in} g_{\rm out}
\label{tod}
\ee
up to some trivial dynamics, defined by a function
analytic in $G=D_0\cup D_\infty$ ($g=g_{\rm in},\;
\lambda\in D_0$; $g=g_{\rm out},\;
\lambda\in D_\infty$).

As we know, {\em arbitrary} function on the
circle having no zeroes (arbitrary loop)
can be represented in the form (\ref{tod}).
So in the case of Toda lattice hierarchy the
objects of the theory nontrivially depend
on the whole loop group (i.e. on discrete time,
positive and negative continuous times),
$\tau$-function in this case is a functional
of arbitrary $g$ on the circle. Discrete
dynamics is correctly defined, because in this case
$\lambda^n$ is a function having in $G$
equal number of zeroes and poles.

The explicit parametrization of the loop group element
$g$ in terms of times is given by
\be
g(n,{\bf x},\nu)=g(n,{\bf x_+},{\bf x_-},\nu)=
\nu^n\exp\left(\sum_{k=1}^{\infty}x_{(+)k}\nu^{-k}  +
x_{(-)k}\nu^{k}\right).
\ee
The compact form of 2D Toda lattice hierarchy
in terms of the function $\chi(\lambda,\mu,k,{\bf x})$
can be obtained from (\ref{HIROTA}) using
special combinations of group elements
$g_1 g_2^{-1}$ with simple analytical properties.
To define dynamics with respect to continuous times,
let us take
\bea
&&g_0(\nu):=g_1(\nu)g_2^{-1}(\nu)=
{\nu-a_+\over\nu-b_+}
{\nu-a_-\over\nu-b_-},\nn\\
&&a_+,b_+\in D_{0},\;a_-,b_-\in D_{\infty}, \nn\\
&&g_0(\nu)=g_+\times (g_-)^{-1},\quad
g_+={\nu-a_+\over\nu-b_+},\quad
g_-={\nu-b_-\over\nu-a_-}.
\eea
Substituting the expression for $g_1 g_2^{-1}$ into the
generalized Hirota bilinear identity (\ref{HIROTA}),
one obtains
\begin{eqnarray}
&&g_0(\mu)\chi(\lambda,\mu,g\times g_+)-
g_0(\lambda)\chi(\lambda,\mu,g\times g_-)+
\nonumber \\
&&
{\rm Res}(g_0)|_{b_+}\;
\chi(\lambda,b_+,g\times g_+)
\chi(b_+,\mu,g\times g_-)+
\nn\\
&&
{\rm Res}(g_0)|_{b_-}\;
\chi(\lambda,b_-,g\times g_+)
\chi(b_-,\mu,g\times g_-)
=0.
\label{TDbasic1}
\end{eqnarray}
The equations of Toda lattice hierarchy are generated
by equation (\ref{TDbasic1}) with
$b_+=0$, $b_-=\infty$, or, in terms of $g_0$
\bea
g_0(\nu)=g_1(\nu)g_2^{-1}(\nu)=(1-{\nu\over a_-})(1-{a_+\over\nu})=
\exp\left(\sum_{i=1}^{\infty}
{1\over i }(( {\nu\over a_-})^i+ ({a_+\over \nu})^{i})\right),\\
|a_+|<1, |a_-|>1\nn\,.
\eea
Substituting into equation (\ref{TDbasic1})
parametrization of loop
group elements in terms of times, we obtain
\begin{eqnarray}
&&
(1-{\mu\over a_-})(1-{a_+\over\mu})
\chi(\lambda,\mu,n,{\bf x_+}+[{\bf a_+}],{\bf x_-})-
\nn\\
&&
(1-{\lambda\over a_-})(1-{a_+\over\lambda})
\chi(\lambda,\mu,n,{\bf x_+},{\bf x_-}-[{\bf a_-}])+
\nonumber \\
&&
-a_+\;
\chi(\lambda,0,n,{\bf x_+}+[{\bf a_+}],{\bf x_-})
\chi(0,\mu,n,{\bf x_+},{\bf x_-}-[{\bf a_-}])
\nn\\
&&
-{1\over a_-}\;
\chi(\lambda,\infty,n,{\bf x_+}+[{\bf a_+}],{\bf x_-})
\chi(\infty,\mu,n,{\bf x_+},{\bf x_-}-[{\bf a_-}])
=0,
\label{TDbasic2}
\end{eqnarray}
where $\chi(\lambda,\infty)=\chi(\lambda,\mu)\times\mu|_{\mu=\infty}$.
In terms of CBA and BA functions $\psi(\lambda,\mu)$ and
$\psi_0(\lambda)$, $\tilde\psi_0(\mu)$
$\psi_\infty(\lambda)$, $\tilde\psi_\infty(\mu)$ we have
\begin{eqnarray}
&&
\psi(\lambda,\mu,n,{\bf x_+}+[{\bf a_+}],{\bf x_-})
-
\psi(\lambda,\mu,n,{\bf x_+},{\bf x_-}-[{\bf a_-}])=
\nonumber \\
&&
a_+
\psi_0(\lambda,n,{\bf x_+}+[{\bf a_+}],{\bf x_-})
\tilde\psi_0(\mu,n,{\bf x_+},{\bf x_-}-[{\bf a_-}])+
\nn\\
&&
{1\over a_-}
\psi_{\infty}(\lambda,n,{\bf x_+}+[{\bf a_+}],{\bf x_-})
\tilde\psi_\infty(\mu,n,{\bf x_+},{\bf x_-}-[{\bf a_-}]).
\label{Todabasic}
\end{eqnarray}
To get equations defining the discrete dynamics, we take
\be
g_0(\nu)=g_1(\nu)g_2^{-1}(\nu)={1\over \lambda}
\ee
or
\be
g_0(\nu)=g_1(\nu)g_2^{-1}(\nu)=\lambda.
\ee
In the first case the equation for the function
$\chi$ reads
\begin{eqnarray}
&&
{1\over\mu}
\chi(\lambda,\mu,n+1,{\bf x_+},{\bf x_-})
-{1\over\lambda}
\chi(\lambda,\mu,n,{\bf x_+},{\bf x_-})=
\nonumber \\
&&
\chi(\lambda,0,n+1,{\bf x_+},{\bf x_-})
\chi(0,\mu,n,{\bf x_+},{\bf x_-}),
\label{TDbasic3}
\end{eqnarray}
and in the second case
\begin{eqnarray}
&&
{\mu}
\chi(\lambda,\mu,n,{\bf x_+},{\bf x_-})
-{\lambda}
\chi(\lambda,\mu,n+1,{\bf x_+},{\bf x_-})=
\nonumber \\
&&
\chi(\lambda,\infty,n,{\bf x_+},{\bf x_-})
\chi(\infty,\mu,n+1,{\bf x_+},{\bf x_-}).
\label{TDbasic4}
\end{eqnarray}
In terms of CBA and BA functions one gets
\begin{eqnarray}
&&
\psi(\lambda,\mu,n+1,{\bf x_+},{\bf x_-})
-
\psi(\lambda,\mu,n,{\bf x_+},{\bf x_-})=
\nonumber \\
&&
\psi_0(\lambda,n+1,{\bf x_+},{\bf x_-})
\tilde\psi_0(\mu,n,{\bf x_+},{\bf x_-})
\nonumber\\
&&
\psi(\lambda,\mu,n,{\bf x_+},{\bf x_-})
-
\psi(\lambda,\mu,n+1,{\bf x_+},{\bf x_-})=
\nonumber \\
&&
\psi_\infty(\lambda,n,{\bf x_+},{\bf x_-})
\tilde\psi(\mu,n+1,{\bf x_+},{\bf x_-}).
\label{TodaD}
\end{eqnarray}
Equations (\ref{Todabasic}), (\ref{TodaD}) define the generalized
2D Toda lattice hierarchy. The discrete dynamics can be resolved in
both directions explicitly (i.e. $\chi(n+1)$, $\chi(n-1)$ can be
expressed through $\chi(n)$) using the determinant formula
\cite{NLS}.
\section{Toda lattice hierarchy: 2DTL, 2D Volterra chain,
2DTL singularity manifold equation}
At this point we would like to derive the simplest equations
of generalized 2D Toda lattice hierarchy. They contain
three variables: one discrete variable $n$ and two
continuous variables $x_{(+)1}$, $x_{(-)1}$
(we will denote them $x_+$ and $x_-$). We
will derive three different equation corresponding
to different levels of the tower of equations
of generalized hierarchy: the main equation
(the standard two-dimensional Toda lattice
equation), the modified equation (i.e. equation
for Baker-Akhiezer function) and singularity
manifold type equation (arising here as an equation
for the CBA function). It is interesting to
note that the second and the third equation
(not to mention the first)
are already known in literature in different
contexts (see \cite{Shabat}, \cite{Fer},
\cite{Gibbon}).

The basic equations for our derivation are
equations (\ref{TDbasic3}), (\ref{TDbasic4})
and the zero order terms of expansion of
(\ref{TDbasic2}) in $a_+$, $1/a_-$, i.e.
\begin{eqnarray}
&&
({\partial~~~\over\partial x_+}+{1\over\lambda}-{1\over\mu})
\chi(\lambda,\mu,n,{\bf x_+},{\bf x_-})
=
\nonumber \\
&&
\chi(\lambda,0,n,{\bf x_+},{\bf x_-})
\chi(0,\mu,n,{\bf x_+},{\bf x_-}),
\nn\\
&&
({\partial~~~\over\partial x_-}+{\lambda}-{\mu})
\chi(\lambda,\mu,n,{\bf x_+},{\bf x_-})
=
\nonumber \\
&&
\chi(\lambda,\infty,n,{\bf x_+},{\bf x_-})
\chi(\infty,\mu,n,{\bf x_+},{\bf x_-}).
\label{TD1}
\end{eqnarray}
In terms of CBA and BA functions instead of
(\ref{TD1}) we have
\begin{eqnarray}
&&
{\partial~~~\over\partial x_+}
\psi(\lambda,\mu,n,{\bf x_+},{\bf x_-})
=
\psi_0(\lambda,n,{\bf x_+},{\bf x_-})
\tilde\psi_0(\mu,n,{\bf x_+},{\bf x_-}),
\nn\\
&&
{\partial~~~\over\partial x_-}
\psi(\lambda,\mu,n,{\bf x_+},{\bf x_-})
=
\psi_\infty(\lambda,n,{\bf x_+},{\bf x_-})
\tilde\psi_\infty(\mu,n,{\bf x_+},{\bf x_-}).
\label{TD2}
\end{eqnarray}
Combining (\ref{TDbasic3}), (\ref{TDbasic4})
and (\ref{TD1}), one obtains
\begin{eqnarray}
&&
({\partial~~~\over\partial x_+}+{1\over\lambda})
\chi(\lambda,\mu,n,{\bf x_+},{\bf x_-})-
{1\over\lambda}
\chi(\lambda,\mu,n-1,{\bf x_+},{\bf x_-})
=
\nonumber \\
&&
\chi(\lambda,0,n,{\bf x_+},{\bf x_-})(
\chi(0,\mu,n,{\bf x_+},{\bf x_-})-
\chi(0,\mu,n-1,{\bf x_+},{\bf x_-})),
\nn\\
&&
({\partial~~~\over\partial x_-}+{\lambda})
\chi(\lambda,\mu,n,{\bf x_+},{\bf x_-})
-{\lambda}
\chi(\lambda,\mu,n+1,{\bf x_+},{\bf x_-})
=
\nonumber \\
&&
\chi(\lambda,\infty,n,{\bf x_+},{\bf x_-})(
\chi(\infty,\mu,n,{\bf x_+},{\bf x_-})-
\chi(\infty,\mu,n+1,{\bf x_+},{\bf x_-})).
\label{TD3}
\end{eqnarray}
Taking the first equation at $\lambda=\infty$,
$\mu=0$, the second at $\lambda=0$,
$\mu=\infty$, one obtains
the following equations
\bea
&&
\partial_+u(n)=
(U(n)-U(n-1))
u(n),
\nn\\&&
\partial_-v(n)=
(V(n)-V(n+1))
v(n).
\label{TE1}
\eea
Here we have introduced the notations
\bea
&&
u(n)=\chi(\infty,0,n,{\bf x_+},{\bf x_-}),
\nn\\&&
v(n)=\chi(0, \infty, n,{\bf x_+},{\bf x_-}),
\nn\\&&
U(n)=\chi_{\rm r}(0, 0, n,{\bf x_+},{\bf x_-}),
\nn\\&&
V(n)=\chi_{\rm r}(\infty,\infty, n,{\bf x_+},{\bf x_-}),
\nn
\eea
and we omit arguments ${\bf x_+},{\bf x_-}$
common for all functions.

Equations (\ref{TD1}) taken respectively
at $\lambda=\infty$,
$\mu=\infty$; $\lambda=0$, $\mu=0$ give rise to
the following relations
\bea
&&
\partial_+V(n)=u(n)v(n)
-1,
\nn\\&&
\partial_-U(n)=u(n)v(n)
-1,
\label{TE2}
\eea
which together with (\ref{TE1}) form a closed system
of equations.

There is also an additional relation implied by
(\ref{TDbasic3}), (\ref{TDbasic4}):
\be
u(n+1)v(n)=1.
\label{TE3}
\ee
This formula shows that the functions
$u(n)$
and $v(n)$
are not independent.

The system of equations (\ref{TE1}), (\ref{TE2}),
(\ref{TE3}) represent some special form of
classical 2D Toda lattice equations. To show it,
let us derive more standard forms of Toda
lattice from this system.  First, it is easy
to show that
\bea
&&
\partial_+\partial_-\ln u(n)=u(n)v(n)-u(n-1)v(n-1),
\nn\\&&
\partial_+\partial_-\ln v(n)=u(n)v(n)-u(n+1)v(n+1).
\eea
Thus for the function $h(n)=u(n)v(n)$
we have an equation
\be
\partial_+\partial_-\ln h(n)=2h(n)-h(n-1)-h(n+1),
\ee
which is 2D Toda lattice equation in the Darboux form.

Since, according to (\ref{TE3}), the functions $u(n)$
and $v(n)$ are not independent, we can write
an equation for one function
\be
\partial_+\partial_-\ln u(n)={u(n)\over u(n+1)}-
{u(n-1)\over u(n)}.
\ee
In terms of $\theta(n)=\ln u(n)$ we have
\be
\partial_+\partial_-\theta(n)=\exp(\theta(n)-\theta(n+1))-
\exp(\theta(n-1)-\theta(n)).
\ee
This is just a standard form of the 2D Toda lattice
equation. An original bilinear form,
introduced by Toda, follows from
addition formulae for the $\tau$-function,
which will be given in the next section.

Now we proceed to the modified Toda lattice
equation. In this case we will use the
equations for generic wave functions,
which are given by
(\ref{TodaD})
and (\ref{TD2}) integrated over $\lambda$ and $\mu$
with some weight functions $\rho(\lambda)$
and $\eta(\mu)$ (in fact, each of them in matrix
notations is a pair of functions defined on
`in' unit circle and `out' unit circle, see also
(\ref{Phi})):
\begin{eqnarray}
&&
\Phi(n
)-
\Phi(n+1
)=
\Psi_0(n+1
)
\widetilde\Psi_0(n
),
\nonumber\\
&&
\Phi(n
)
-
\Phi(n-1
)=
\Psi_\infty(n-1
)
\widetilde\Psi_\infty(n
)
\label{TodaD'}
\end{eqnarray}
(we have omitted common arguments
${\bf x_+},{\bf x_-}$)

Combining these equations, one
obtains a linear problem for the
modified Toda lattice equation
\bea
&&
\partial_+ \Phi=W_+(1-T)\Phi, \quad W_+={\psi_0\over T\psi_0},
\nn\\&&
\partial_- \Phi=W_-(1-T^{-1})\Phi, \quad
W_-={\psi_\infty\over T^{-1}\psi_\infty}.
\eea
The compatibility condition gives the modified Toda lattice
equations
\bea
&&\partial_-W_+=W_+(1-T)W_-,
\nn\\&&
\partial_+W_-=W_-(1-T^{-1})W_+.
\label{MT}
\eea
These equations quite recently appeared in literature
\cite{Shabat}, \cite{Fer}
under the name of two-dimensional Volterra chain.
It is interesting to note that in \cite{Fer}
they arose in the geometrical context.

Proceeding further, one can obtain an equation
for the function $\Phi$, using linear operators of
modified Toda lattice to express the functions
$W_+$, $W_-$ via the function $\Phi$ and then
substituting the final result to equations
(\ref{MT}). Both equations give rise to the same
equation
\be
\partial_+\partial_-\Phi+
\partial_+\Phi\cdot \partial_-\Phi
\left({1\over T\Phi-\Phi}+{1\over T^{-1}\Phi-\Phi}
\right)=0.
\label{Tsingman}
\ee
This equation was also introduced in different
contexts in \cite{Shabat}, \cite{Fer}. Another
interesting feature of equation (\ref{Tsingman})
could be its connection with the Painleve analysis of
the 2DTL. Making comparison with one-dimensional
Toda lattice case \cite{Gibbon},
and recalling analogous equation for the
scalar KP case, it is natural to suggest that the
equation (\ref{Tsingman}) represent a singularity manifold equation
for the 2DTL.

Equation (\ref{Tsingman}) generates 2DTL and two-dimensional
Volterra chain in different ways: first, in terms of
Combescure invariants,  and second,
in terms of expansion in $\lambda$, $\mu$ of the
function $\psi(\lambda,\mu,n,{\bf x_+},{\bf x_-})$ which
satisfies equation (\ref{Tsingman}).

The Combescure invariants for (\ref{Tsingman})
are given by the formulae
\bea
&&
W_+=-{\partial_+\Phi\over \Delta\Phi},
\nn\\&&
W_-= {\partial_-T\Phi\over \Delta\Phi},
\label{Combsing}
\eea
(we have written down only `left' invariants).
For two-dimensional Volterra chain
the Combescure invariants look like
\bea
&&
u=W_-\exp(-\Delta^{-1}(W_-+T W_+)),
\nn\\&&
v=W_+\exp(\Delta^{-1}(W_-+T W_+)).
\label{CombVol}
\eea
Substituting (\ref{Combsing}) to (\ref{CombVol}),
one obtains expressions of the full
Combescure group invariants for
singularity manifold equation (\ref{Tsingman}).

At last, we present the formulae for
the hierarchies of two-dimensional
Toda, Volterra and 2D Toda singularity manifold
equations. The derivation is very similar to the case
of DS hierarchy, so we will omit it.

The 2DTL hierarchy can be represented as an
equation
\bea
&&
T_+T_-^{-1}u+{a_+\over a_-}Tu=
T_+uT_-^{-1}u
\left({1\over u}+{a_+\over a_-}{1\over T_+T_-^{-1}T^{-1}u}\right),
\eea
where a convention for shift and difference operators
is standard
\bea
&&
T_+:{\bf x_+}\rightarrow {\bf x_+}+[a_+],
\nn\\&&
T_-:{\bf x_-}\rightarrow {\bf x_+}+[a_-^{-1}],
\nn\\&&
\Delta_+={T_+-1},
\nn\\&&
\Delta_-=(T_--1).
\eea
The 2D Volterra hierarchy reads
\bea
&&
a_-\Delta_-W_++(T_-W_+)(TW_-)-(T_+W_-)W_+=0,
\nn\\&&
1/a_+\Delta_+W_-+(T_+W_-)(T^{-1}W_+)-(T_-W_+)W_-=0,
\eea
where $W_+={T_+\Psi_0\over T\Psi_0}$,
$W_-={T_-\Psi_\infty\over T^{-1}\Psi_\infty}$.

Finally, 2DTL singularity manifold equation
hierarchy looks like
\bea
&&
\{T_-(T_+-1)\Phi\}\{T(T_--T^{-1})\Phi\}\{T_+(1-T^{-1})\Phi\}=
\nn\\&&
\{(T_+-1)\Phi\}\{T_+(T_--T^{-1})\Phi\}\{T_-(T-1)\Phi\}.
\eea
This equation possesses a symmetry $T_+\rightarrow T_-$,
$T_-\rightarrow T_+$, $T\rightarrow T^{-1}$.
Though this symmetry is not obvious but it can be easily verified.
\section{Addition formulae for Toda lattice hierarchy}
The $\tau$-functional substitution
\bea
\chi(\lambda,\mu,g)={1\over\lambda-\mu}{\tau(g_{(\lambda,\mu)}g)
\over
\tau(g)},
\\
g_{(\lambda,\mu)}={\lambda-\nu\over\mu-\nu},
\nn
\eea
to (\ref{TDbasic1}),
(\ref{TodaD})
leads to addition formulae for the $\tau$-function,
the expansion of which gives the equations of the
Toda lattice hierarchy in terms of the $\tau$-function.

The simplest addition formulae follow from (\ref{TodaD})
\begin{eqnarray}
&&{1\over \mu(\lambda-\mu)}
\tau(g\times g_{\rm d}g_{(\lambda,\mu)})
\tau(g)-
{1\over\lambda(\lambda-\mu)}\tau(g\times g_{(\lambda,\mu)})
\tau(g\times g_d)+
\nonumber \\
&&
{1\over\lambda\mu}\tau(g\times g_{\rm d}g_{(\lambda,0)})
\tau(g\times g_{(0,\mu)})=0.
\label{TDaddition}
\end{eqnarray}

This is a functional form of addition
formulae, containing four components characterized
by the position of $\lambda$, $\mu$ with respect
to the unit circle. Using parametrization
of loop group elements in terms of times,
we obtain a standard form of addition formulae.
It is convenient to write down first the factorization
formulae for the function $g_{(\lambda,\mu)}$,
representing it as a product of functions analytic
inside and outside the unit circle for different
positions of $\lambda$ and $\mu$. For $\lambda$,
$\mu$ both belonging to $D_0$  ($D_\infty$) the
factorization is trivial, because  $g_{(\lambda,\mu)}$
is analytic outside (inside) the unit circle.
For $\lambda$, $\mu$ in different domains it is
given by
\bea
&&g_{(\lambda_+,\mu_-)}=g_{\rm d} g_{(\lambda_+,0)}g_{(\infty,\mu_-)},
\nn\\
&&g_{(\lambda_-,\mu_+)}=g_{\rm d}^{-1} g_{(\lambda_-,\infty)}g_{(0,\mu_+)},
\nn\\
&&g_{\rm d}(\nu)=\nu,
\eea
here the subscript +, - just indicates the position inside or
outside the unit circle.

In terms of time variables, the action
of loop group elements represented in the formula
(\ref{TDaddition}) (taking into account the factorization
formulae) looks like
\bea
&&g_{(\lambda_+,\mu_+)}\rightarrow  {\bf x_+}+[\mu_+]-[\lambda_+],
\nn\\
&&g_{(\lambda_-,\mu_-)}\rightarrow  {\bf x_-}+[\mu_-^{-1}]-[\lambda_-^{-1}],
\nn\\
&&g_{(\lambda_+,\mu_-)}\rightarrow n+1,{\bf x_+}-[\lambda_+],
{\bf x_-}+[\mu_-^{-1}],
\nn\\
&&g_{(\lambda_-,\mu_+)}\rightarrow n-1,{\bf x_+}+[\mu_+],
{\bf x_-}-[\lambda_-^{-1}].
\eea
Substituting these formulae to
(\ref{TDaddition}), we obtain a standard form of addition
formulae for the Toda lattice hierarchy. It occurs that
off-diagonal components of (\ref{TDaddition}) give
the same formula with interchanged $\lambda$, $\mu$,
so we have three addition formulae
\bea
&&\lambda\tau_n^{(x_++[\mu],x_-)}\tau_n^{(x_+,x_-+[\lambda^{-1}])}-
\mu\tau_{n-1}^{(x_++[\mu],x_-)}\tau_{n+1}^{(x_+,x_-+[\lambda^{-1}])}=
\nn\\
&&(\lambda-\mu)\tau_n^{(x_+,x_-)}\tau_n^{(x_++[\mu],x_-+[\lambda^{-1}])},
\nn\\
&&|\lambda|>1,|\mu|<1,
\label{TAA}
\\
&&\lambda\tau_{n+1}^{(x_++[\mu],x_-)}\tau_n^{(x_++[\lambda],x_-)}-
\mu\tau_n^{(x_++[\mu],x_-)}\tau_{n+1}^{(x_++[\lambda],x_-)}=
\nn\\
&&(\lambda-\mu)\tau_{n+1}^{(x_+,x_-)}\tau_n^{(x_++[\mu]+[\lambda],x_-)},
\nn\\
&&|\lambda|<1,|\mu|<1,
\label{TAB}
\\
&&
\lambda\tau_n^{(x_+,x_-+[\mu^{-1}])}\tau_{n-1}^{(x_+,x_-+[\lambda^{-1}])}-
\mu\tau_{n-1}^{(x_+,x_-+[\mu^{-1}])}\tau_n^{(x_+,x_-+[\lambda^{-1}])}=
\nn\\&&
(\lambda-\mu)\tau_{n-1}^{(x_+,x_-)}
\tau_n^{(x_+,x_-+[\mu^{-1}]+[\lambda^{-1}])},
\nn\\&&
|\lambda|>1,|\mu|>1.
\label{TAC}
\eea
In fact, we have used only the first equation of the pair
of equations  (\ref{TodaD}), but the second equation gives exactly
the same set of addition formulae.

The formula (\ref{TAA}) is a well-known simplest addition formula
for the Toda lattice hierarchy. The formulae  (\ref{TAB}), (\ref{TAC})
do not
shift negative (positive) times and it is connected with
the B\"acklund transformations for the KP equation.

To get more generic addition formulae, one could use the formula
(\ref{TDbasic1}). To get more symmetric expressions,
we will use a special version of this formula
corresponding to the choice
\bea
&&g_0(\nu):=g_1(\nu)g_2^{-1}(\nu)=
{\nu-\xi\over\nu-\eta},
\eea
where we do not specify from the beginning the position
of the points $\lambda_1$, $\mu_1$. This choice leads to the following
functional form of the addition formula
\begin{eqnarray}
&&(\mu-\eta)(\lambda-\xi)
\tau(gg_{(\xi,\eta)}g_{(\lambda,\mu)})
\tau(g)-
(\lambda-\eta)(\mu-\xi)\tau(gg_{(\lambda,\mu)})
\tau(gg_{(\xi,\eta)})+
\nonumber \\
&&
(\lambda-\mu)
(\eta-\xi)
\tau(gg_{(\xi,\mu)})
\tau(gg_{(\lambda,\eta)})=0.
\label{TA0}
\end{eqnarray}
The formula (\ref{TA0})
contains 16 components depending on the position of
four parameters with respect to unit circle.
It is convenient to introduce
the function showing the position of each parameter
\bea
i_{\lambda}=\left\{
\begin{array}{ccc}
{1\over2}&\quad&|\lambda|<1\\
-{1\over2}&\quad&|\lambda|>1
\end{array}
\right.
\eea
Then the addition formulae take the form
\begin{eqnarray}
&&(\mu-\eta)(\lambda-\xi)
\tau_{n-i_{\eta}-i_{\mu}}^{({\bf x}+[[\eta]]+[[\mu]])}
\tau_{n-i_{\lambda}-i{\xi}}^{({\bf x}+[[\lambda]]+[[\xi]])}-
\nn\\&&
(\lambda-\eta)(\mu-\xi)
\tau_{n-i_{\mu}-i_{\xi}}^{({\bf x}+[[\xi]]+[[\mu]])}
\tau_{n-i_{\lambda}-i_{\eta}}^{({\bf x}+[[\lambda]]+[[\eta]])}+
\nonumber \\
&&
(\lambda-\mu)
(\eta-\xi)
\tau_{n-i_{\lambda}-i_{\mu}}^{({\bf x}+[[\mu]]+[[\lambda]])}
\tau_{n-i_{\eta}-i_{\xi}}^{({\bf x}+[[\eta]]+[[\xi]])}=0,
\label{TA01}
\end{eqnarray}
where
\bea
({\bf x}+[[\lambda]])=
\left\{
\begin{array}{lcl}
(x_+ +[\lambda],x_-)&\quad&|\lambda|<1\\
(x_+,x_-+[\lambda^{-1}])&\quad&|\lambda|>1
\end{array}
\right.
\nn
\eea
Due to its symmetric form, the formula (\ref{TA01})
contains only five different addition formulae
corresponding to all four parameters inside the unit circle,
one parameter outside, two parameters outside,
three parameters outside, all four parameters outside.
The first and the last case just give the KP addition formulae,
so we will consider only remaining three cases.
First we write down the addition
formulae for the case when two variables
(say, $\lambda$ and $\mu$) are inside the unit circle:
\begin{eqnarray}
&&(\mu-\eta)(\lambda-\xi)
\tau_{n}^{({\bf x_+}+[\mu],{\bf x_-}+[\eta^{-1}])}
\tau_{n}^{({\bf x_+}+[\lambda],{\bf x_-}+[\xi^{-1}])}-
\nn\\&&
(\lambda-\eta)(\mu-\xi)
\tau_{n}^{({\bf x_+}+[\mu],{\bf x_-}+[\xi^{-1}])}
\tau_{n}^{({\bf x_+}+[\lambda],{\bf x_-}+[\eta^{-1}])}+
\nonumber \\
&&
(\lambda-\mu)
(\eta-\xi)
\tau_{n-1}^{({\bf x_+}+[\mu]+[\lambda],{\bf x_-})}
\tau_{n+1}^{({\bf x_+},{\bf x_-}+[\eta^{-1}]+[\xi^{-1}])}=0.
\label{TA0A}
\eea
The structure of this addition formula is similar
to the formula (\ref{TAA}), and in fact  (\ref{TA0A})
is a generalized version of (\ref{TAA}).

When one parameter (say, $\lambda$) is inside the unit
circle, we obtain the following addition formula
\begin{eqnarray}
&&(\mu-\eta)(\lambda-\xi)
\tau_{n+1}^{({\bf x_+},{\bf x_-}+[\eta^{-1}]+[\mu^{-1}])}
\tau_{n}^{({\bf x_+}+[\lambda],{\bf x_-}+[\xi^{-1}])}-
\nn\\&&
(\lambda-\eta)(\mu-\xi)
\tau_{n+1}^{({\bf x_+},{\bf x_-}+[\xi^{-1}]+[\mu^{-1}])}
\tau_{n}^{({\bf x_+}+[\lambda],{\bf x_-}+[\eta^{-1}])}+
\nonumber \\
&&
(\lambda-\mu)
(\eta-\xi)
\tau_{n}^{({\bf x_+}+[\lambda],{\bf x_-}+[\mu^{-1}])}
\tau_{n+1}^{({\bf x_+},{\bf x_-}+[\eta^{-1}]+[\xi^{-1}])}=0.
\label{TA0B}
\end{eqnarray}
And the last case ($\lambda$ outside the unit circle)
gives
\begin{eqnarray}
&&(\mu-\eta)(\lambda-\xi)
\tau_{n-1}^{({\bf x_+}+[\eta]+[\mu],{\bf x_-})}
\tau_{n}^{({\bf x_+}+[\xi],{\bf x_-}+[\lambda^{-1}])}-
\nn\\&&
(\lambda-\eta)(\mu-\xi)
\tau_{n-1}^{({\bf x_+}+[\xi]+[\mu],{\bf x_-})}
\tau_{n}^{({\bf x_+}+[\eta],{\bf x_-}+[\lambda^{-1}])}+
\nonumber \\
&&
(\lambda-\mu)
(\eta-\xi)
\tau_{n}^{({\bf x_+}+[\mu],{\bf x_-}+[\lambda^{-1}])}
\tau_{n-1}^{({\bf x_+}+[\eta]+[\xi],{\bf x_-})}=0.
\label{TA0C}
\end{eqnarray}
The formulae (\ref{TA0B}), (\ref{TA0C}) generalize the formulae
(\ref{TAB}), (\ref{TAC}).
\subsection*{Acknowledgments}
The first author (LB) is grateful to the
Dipartimento di Fisica dell'Universit\`a
and Sezione INFN, Lecce, for hospitality and support;
(LB) also acknowledges partial support from the
Russian Foundation for Basic Research under grant
No 96-01-00841 and by INTAS under grant 93-166.

\end{document}